\documentclass[twocolumn]{aa}
\usepackage{txfonts}
\usepackage{graphicx}
\usepackage{epstopdf}

\usepackage{natbib}
\bibpunct{(}{)}{;}{a}{}{,}

\def\ie{\emph{i.e.}}

\def\deg{\ifmmode^\circ\else$^\circ$\fi}

\def\ah{\ifmmode{^\textrm{\scriptsize h}}\else{$^\textrm{\scriptsize h}$}\fi}
\def\am{\ifmmode{^\textrm{\scriptsize m}}\else{$^\textrm{\scriptsize m}$}\fi}
\def\as{\ifmmode{^\textrm{\scriptsize s}}\else{$^\textrm{\scriptsize s}$}\fi}
\def\alphaTF{\ifmmode{\alpha_{\mathrm{\,{\small TF}}}}\else{$\alpha_{\mathrm{\,{\small TF}}}$}\fi}
\def\etaBpri{\ifmmode{\eta_{\mathrm{B1}}}\else{$\eta_{\mathrm{B1}}$}\fi}
\def\etaDpri{\ifmmode{\eta_{\mathrm{D1}}}\else{$\eta_{\mathrm{D1}}$}\fi}
\def\etaBsec{\ifmmode{\eta_{\mathrm{B2}}}\else{$\eta_{\mathrm{B2}}$}\fi}
\def\etaDsec{\ifmmode{\eta_{\mathrm{D2}}}\else{$\eta_{\mathrm{D2}}$}\fi}

\begin{document}

\title{Growth of galactic bulges by mergers}

\subtitle{II. Low-density satellites}
\author{M.~Carmen Eliche-Moral\inst{1}, Marc Balcells\inst{1}, J.~Alfonso L.~Aguerri\inst{1} \& A.~C\'esar 
Gonz\'alez-Garc\'{\i}a\inst{1}}
\offprints{\email{mcem@iac.es; balcells@iac.es; jalfonso@iac.es; cglez@iac.es}}

\institute{Instituto de Astrof\'{\i}sica de Canarias, C/ V\'\i a
L\'actea, E-38200 La Laguna, Tenerife, Spain}
\date{Received April 4th 2006 / Accepted June 22th 2006}

\abstract
{Satellite accretion events have been invoked for mimicking the internal secular evolutionary processes of bulge growth. However, N-body simulations of satellite accretions have paid little attention to the evolution of bulge photometric parameters, to the processes driving this evolution, and to the consistency of this evolution with observations. 
}
{We want to investigate whether satellite accretions indeed drive the growth of bulges, and whether they are consistent with global scaling relations of bulges and discs.  
}
{We perform $N$-body models of the accretion of satellites onto disc galaxies. 
A Tully-Fisher ($M\propto V_{\rm rot}^\alphaTF$) scaling between primary and satellite ensures that density ratios, critical to the outcome of the accretion, are realistic.  We carry out a full structural, kinematic and dynamical analysis of the evolution of the bulge mass, bulge central concentration, and bulge-to-disc scaling relations.}
{
The remnants of the accretion have bulge-disc structure.  
Both the bulge-to-disc ratio ($B/D$) and the S\'ersic index (\emph{n}) of the remnant bulge increase as a result of the accretion, with moderate final bulge S\'ersic indices: $n = 1.0$ to 1.9.  Bulge growth occurs no matter the fate of the secondary, which fully disrupts for $\alphaTF=3$ and partially survives to the remnant center for  \alphaTF = 3.5 or 4.  
Global structural parameters evolve following trends similar to observations. We show that the dominant mechanism for bulge growth is the inward flow of material from the disc to the bulge region during the satellite decay.}
{The models confirm that the growth of the bulge out of disc material, a central ingredient of secular evolution models, may be triggered externally through satellite accretion.}

\authorrunning{M.~C.~Eliche-Moral et al.}

\titlerunning{Galactic bulges and mergers. II.}
\maketitle

\section{Introduction}
\label{Sec:Introduction} 

Both observationally and theoretically, the central bulges of disc galaxies have received much attention since the last decade \citep[see reviews by ][]{Wyse97,Kormendy04}.  However, we are far from understanding the processes that led to their formation.  Several scenarios have been proposed.  The similarities between bulges and ellipticals have prompted models of fast, early formation from the collapse of a protogalactic halo \citep{Eggen62,Gilmore98}, or as the result of violent interactions and major mergers of collapsed pre-galactic clumps prior to the formation of the thin disc \citep[][]{Kauffmann93,Kauffmann94,Kauffmann96}. But the realization that the \emph{bulges are old} paradigm is not entirely true \citep[][]{Kormendy93} has led to the development of models in which bulges grow over long timescales, after the formation of the disc. Processes that take many galactic rotation periods to complete \citep[of the order of several Gyr, see KK04; ][]{Athanassoula05} have been termed \emph{secular evolution} processes.  
They include bulge growth from an internal disc bar instability followed either by a vertical instability in the bar itself \citep{Hohl71,Combes81,Pfenniger90, Debattista04}, or by bar dissolution due to a central mass concentration \citep{Hasan90,Hasan93}. KK04 list  satellite accretion as a secular process as well, triggered by the environment of the galaxy rather than internally.  Bulge formation/growth could come from the accreted satellites themselves, from the disc \citep{Pfenniger93,Mihos95}, or from a combination of both \citep{Aguerri01}.  


Minor mergers have been extensively studied using numerical simulations in the past \citep[][to name just a few]{Quinn93,Velazquez99,Bournaud04,Bournaud05,
Aceves05a,Aceves05b,Aceves06}. However, interest in satellite accretion has focussed more on the effects on the disc, such as thickening and warping \citep{Walker96,Huang97,Helmi99}, with little attention given to the evolution of the bulge photometric parameters and to the processes that drive this evolution.  Simulations can easily tell us whether minor mergers can indeed drive the growth of early-type bulges.  They can also tell us whether infall-driven bulge growth is consistent with global scaling relations of bulges and discs.  

ABP01 investigated the previous questions with N-body simulations of the accretion of dense, spheroidal satellites. After the accretion of the satellite, bulge-disc decompositions showed that the bulge grew and the S\'ersic index $n$ increased proportionally to the satellite mass. However, most ABP01 satellites were dense spheroids, not representative of the majority of satellites of disc galaxies in the local Universe; those satellites reached the center of the remnant with little tidal stripping, yielding bulge growth measures that may have been unrealistic.  Accretion-driven bulge growth is sensitive to the density contrast between the primary and the merging satellite.  While high-density satellites such as those of the ABP01 study deposit most of their mass, orbital energy and angular momentum in the center of the remnant, low-density satellites disrupt during the merger, and their mass gets deposited at intermediate radii.  The fate of satellites with realistic densities cannot be ascertained a priori or with analytical means, but it can with simulations.

We have expanded on ABP01 work by running N-body simulations of the accretion of satellites that have lower densities than ABP01 experiments.  Our satellites have more complex inner structure than ABP01 spheroidal satellites: they comprise bulge, disc and halo.  Density ratios between primary and satellite derive from using the Tully-Fisher (TF) relation in the scaling of the primary galaxy and the satellites.  We use our models to study whether satellites with realistic densities drive bulge growth as compared with the ABP01 experiments, and to study what physical processes are involved in the evolution of the central component of the galaxies.  
We will show that the fate of the satellites depends critically on the exponent \alphaTF\ of the TF relation used for density scaling: varying \alphaTF\ over a realistically motivated range the fate varies from survival to full disruption.  
Our models suggest that TF-scaled accretion events produce an increase of both the $B/D$ and the S\'ersic index $n$ of the bulge, as in the high-density models of ABP01,  but the mechanisms that drive these changes are different in both cases. We find that TF-scaled satellite accretion processes alter galaxies systematicaly, giving place to remnants whose properties are scaled between them, depending on the mass ratios and the orbit of the encounter. We will show that the growth of the bulge out of disc material, a central ingredient of secular evolution models,  may be triggered externally through satellite accretion.

We will indistinctly refer to our models as "satellite accretions" and "minor mergers" \citep{Bertschik03,Bertschik04b}.

This paper is organized as follows. Models are described in \S\ref{Sec:models}. 
We explain the bulge-disc decomposition procedure in \S\ref{Sec:profiles}, 
while results on growth of bulges are presented in \S\ref{Sec:growthofbulges} and \S\ref{Sec:whynincreases}. Population mixing is analysed in \S\ref{Sec:populationmixing}, while section \S\ref{Sec:thickening} comments the vertical structure of the final remnants. We analyse phase-space evolution and model kinematics in \S\ref{Sec:dynamics} and \S\ref{Sec:kinematics}, respectively. Scaling relations of discs and bulges in the remnants and the comparison to observed correlations are given in \S\ref{Sec:scaling}. Limitations of the current models are analysed in \S\ref{Sec:limitations}. We discuss the connections between remnants of satellite accretion and pseudobulges  in \S\ref{Sec:discussion}, and finally, we list the main conclusions in \S\ref{Sec:conclusions}. 

\section{Models}
\label{Sec:models}

The primary galaxy model was built using the {\tt GalactICS} code \citep{Kuijken95}.  The model comprises an exponential disc 
\citep{Shu69}, a King bulge \citep{King66}, and a dark halo built as an Evans model \citep{Kuijken94}. The core radius of the King model is 0.15 (units given below), and the concentration parameter is 6.7. The disc extends to 5 length units.  It has an exponential surface density profile both radially and vertically; scale length is $h_\mathrm{D}=1.0$, and scale height is $z_\mathrm{D} = 0.1$ \citep[][]{Guthrie92,deGrijs98}.  
The chosen disc velocity dispersion makes the disc warm with a Toomre stability parameter of $Q$=1.7, enough for preventing bar instabilities in the disc when in isolation. Masses, radii and numbers of particles for each component are given in Table~\ref{Tab:models}. Throughout the paper, a gravitational constant of $G$=1 is used, the length unit is the disc scale length, and the total mass of the primary galaxy is equal to 6.44. Our primary galaxy matches the Milky Way when the units of length, velocity and mass are: $R=4.5$ kpc, $v=220$ km s$^{-1}$, $M=5.1\times 10^{10} M_\odot$.  The corresponding time unit is 20.5 Myr.

\begin{table*}
\begin{minipage}[t]{\textwidth}
\caption{Initial parameters of the primary galaxy and the satellites.}
\label{Tab:models}
\centering
\renewcommand{\footnoterule}{}  
\begin{tabular}{cccccccccccccccc}
\hline\hline
 & &\multicolumn{7}{c}{Number of Particles ($/10^3$)}  & & \multicolumn{6}{c}{Primary Galaxy}\\ \cline{3-9}\cline{11-16}
  NP& & Disc1 & Bulge1 & Halo1 & &  Disc2 & Bulge2 & Halo2 & & $\mathcal{M}_{\mathrm{Bul}}$& $\mathcal{M}_{\mathrm{Disc}}$ 
  & $\mathcal{M}_{\mathrm{Dark}}$ & $r_{\mathrm{Bul}}$ & $h_{\mathrm{D}}$ & $z_{\mathrm{D}}$ \\
 (1)& & (2)  & (3)    & (4)  & & (5)   & (6)    & (7) &  & (8)  & (9)  & (10) 
&(11) &(12) &(13)\\\hline
 185 & &40 & 10 & 90 & & 10 & 5 & 30 & &0.42 & 0.82 & 5.20 & 0.195 &
 1.0&0.1\\[0.05cm]\hline\\
\end{tabular}
\begin{minipage}[t]{\textwidth}
\emph{Description of columns}: (1) Total particle number (/$10^3$). (2) Number of primary disc particles.  (3) Number of primary bulge particles. (4) Number of primary halo particles. (5) Number of satellite disc particles. (6) Number of satellite bulge particles. (7) Number of satellite halo particles. (8) Primary bulge mass. (9) Primary disc mass. (10) Primary halo mass. (11) Primary bulge half-mass radius. (12) Primary disc scale length. (13) Primary disc scale height.
\end{minipage}\end{minipage}
\end{table*}

\begin{table*}
\begin{minipage}[t]{\textwidth}
\caption{Orbital and scaling parameters for the merger experiments.}
\label{Tab:orbits}
\centering
\renewcommand{\footnoterule}{}  
\begin{tabular}{cccccccccccc}
\hline\hline
Model Code & $\mathcal{M}_{\mathrm{Sat}}(\mathrm{Lum})/\mathcal{M}_{\mathrm{G}}(\mathrm{Bulge})$ & 
$\mathcal{M}_{\mathrm{Sat}}/\mathcal{M}_{\mathrm{G}}$  &
$\alphaTF$ & $R_{\mathrm{Sat}}/R_{\mathrm{G}}$ & $L_{\mathrm{z,0}}$ & $e$ & 
$R_{\mathrm{peri}}$ & $V_{\mathrm{peri}}$ & $\theta _{1}$ & $t_\mathrm{full\,merger}$ & $t_\mathrm{total}$\\
 (1)&  (2)  & (3)    & (4)  & (5)   & (6)    & (7) &  (8)  & (9) & (10)& (11) & (12)\\\hline
 M2TF4D & 1/2 & 1/6 & 4.0  & 0.40 & 0.4815 & 0.907 & 0.73 & 1.981 & 30   & $\sim$66& 100\\
 M2TF35D& 1/2 & 1/6 & 3.5  & 0.46 & 0.4815 & 0.907 & 0.73 & 1.981 & 30   & $\sim$72& 100\\
 M2TF3D & 1/2 & 1/6 & 3.0  & 0.54 & 0.4815 & 0.907 & 0.73 & 1.981 & 30   & $\sim$80& 100\\
 M3TF4D & 1/3 & 1/9 & 4.0  & 0.33 & 0.3532 & 0.900 & 0.79 & 1.793 & 30   & $\sim$70& 100\\
 M3TF35D& 1/3 & 1/9 & 3.5  & 0.39 & 0.3532 & 0.900 & 0.79 & 1.793 & 30   & $\sim$80& 100 \\
 M3TF3D & 1/3 & 1/9 & 3.0  & 0.48 & 0.3532 & 0.900 & 0.79 & 1.793 & 30   & $\sim$85& 100 \\
 M6D	& 1/6 & 1/18 & 3.5  & 0.28 & 0.1963 & 0.892 & 0.86 & 1.606 & 30  & $\sim$116& 122 \\
 M2R	& 1/2 & 1/6 & 3.5  & 0.46 & 0.4815 & 0.907 & 0.73 & 1.981 & 150  & $\sim$80& 100 \\
 M3R	& 1/3 & 1/9 & 3.5  & 0.39 & 0.3532 & 0.900 & 0.79 & 1.793 & 150  & $\sim$87& 100 \\
 M6R	& 1/6 & 1/18 & 3.5  & 0.28 & 0.1963 & 0.892 & 0.86 & 1.606 & 150 & $\sim$142& 154\\[0.05cm]
\hline\\\end{tabular}
\begin{minipage}[t]{\textwidth}
\emph{Description of columns}: 
(1) Model code. 
(2) Initial mass ratio between luminous satellite material and primary bulge material. 
(3) Initial mass ratio between luminous satellite material and luminous primary material. 
(4) Tully-Fisher index for scaling. 
(5) Ratio of half-mass radii for the luminous distributions of satellite and primary galaxy. (6) Initial orbital angular momentum. (7) Orbital eccentricity. (8) Pericenter distance. (9) Pericenter speed. 
(10) Initial angle between the orbital momentum and the primary disc spin. 
(11) Time of full merger in the units of the simulation. (12) Total time computed in the units of the simulation.
\end{minipage}
\end{minipage}
\end{table*}

The satellite galaxy comprises a dark halo, a disc and a central bulge, built as a scaled replica of the primary.  A physically-motivated size-mass scaling was used, by imposing that the primary and secondary galaxies follow the TF relation. Following  \citet{Gonzalez05}, homologous galaxies with same mass-to-light ratio fulfill $L \propto V^\alphaTF$ whenever sizes $R$ and masses $M$ scale as 

\begin{equation}
R \propto M^{1-2/\alphaTF},
\label{eqn:SizeMassScaling}
\end{equation}
so we apply scaling in eqn.~\ref{eqn:SizeMassScaling} to our models. Central densities scale as

\begin{equation}
\rho \propto M^{\frac{6}{\alphaTF}-2},
\end{equation}
hence, for $M_2/M_1<1$, $\rho_2/\rho_1$ increases as $\alphaTF$ increases, i.e., satellites scaled with $\alphaTF=4.0$ have higher central densities relative to the primary than those with lower \alphaTF. 

Orbital parameters for the merger experiments, satellite masses and half-mass radii are given in Table~\ref{Tab:orbits}. We ran experiments with satellite luminous masses equal to 1:2, 1:3, and 1:6 of the primary bulge mass.  These fractions correspond to luminous mass ratios between the satellite and the primary galaxy of 1:6, 1:9, and 1:18, respectively. 

We use a coordinate system whose origin lies at the system center of mass, the initial merger orbit lies in the $XY$ plane, 
and the satellite lies on the $+X$ axis at the start of the simulation, with orbital angular momentum pointing to $+Z$.  
Orientations of the disc spins are given by standard spherical ($\theta, \phi$) coordinates.  Subindices 1 and 2 denote the primary and satellite galaxies, respectively.  As \citet{Khochfar06} have shown that the halo spin planes and the orbital planes of the mergers between dark matter halos in cosmological simulations do not exhibit any correlation, we have selected arbitrary angles that avoid a perfect spin-orbit coupling in our experiments and that sample direct and retrograde cases. They also indicate that orbits are independent of the progenitor mass and major merger definitions. Therefore, for each satellite mass we ran a direct orbit with $\theta _1 =30^{\circ}$ and a retrograde orbit with $\theta _1 =150^\circ$. In all cases, $\phi _1 =0$\deg, $\theta _2=25$\deg, and $\phi _2=90$\deg. All our experiments exhibit very high eccentricities (see Table\,\ref{Tab:orbits}), according to the cosmological simulations that have found that almost half of all the mergers between dark matter halos are expected to have occurred with $e\sim 1$ \citep{Khochfar06}. Initial orbits were elliptical with apocenter equal to twice the disc outer radius ($\sim$15 length units) and pericenter equal to the disc scale length ($\sim$1 length unit).  Each orbit was run using \alphaTF=3.5.  In addition, for several of the orbit and mass-ratio combinations, we ran models with $\alphaTF = 3.0$ and 4.0, in order to test the response of the system to variations in the density contrast between primary and secondary. 

In the remainder of the paper, models are named with a three-component code: M$m$TF$\alpha$[D/R], where $m$ gives the bulge-to-satellite mass ratio, $\alpha$ denotes the TF exponent (e.g., TF35 for $\alphaTF=3.5$), and "D" or "R" describe the orbit ("D" for direct and "R" for retrograde). Where TF$\alpha$ does not appear, TF35 must be understood for that mass ratio and orbit.

Computations were carried out with SunBlade100 machines. Evolution was computed using the {\small TREECODE\/} of L.~ Hernquist \citep[see][]{Hernquist87,Hernquist90b,Hernquist89,Hernquist90a}, kindly made available by the author. The equations of motion are integrated using a time-centered leapfrog algorithm \citep[][]{Press86} with a variable time step. Gravitational force was softened using a spline kernel, with a constant softening legth $\varepsilon$=0.02. Applying quadrupole-moment corrections with a tolerance parameter $\theta$=0.8, the code computes forces within 1\% of those given by a direct summation.  The primary and the satellite models were allowed to relax separately for about 100 bulge dynamical times prior to placing them in orbit for the merger simulations. Time to full merger ranges from $\sim$65 to $\sim$140 time units. We evolved all models for $\sim$4 halo crossing times beyond full merger to allow the final remnants to reach a near equilibrium state, except models with mass ratio 1:6, which were evolved for $\sim$1-2 halo crossing times.

\begin{figure}[!tbp]
\centering
\includegraphics[width=\columnwidth]{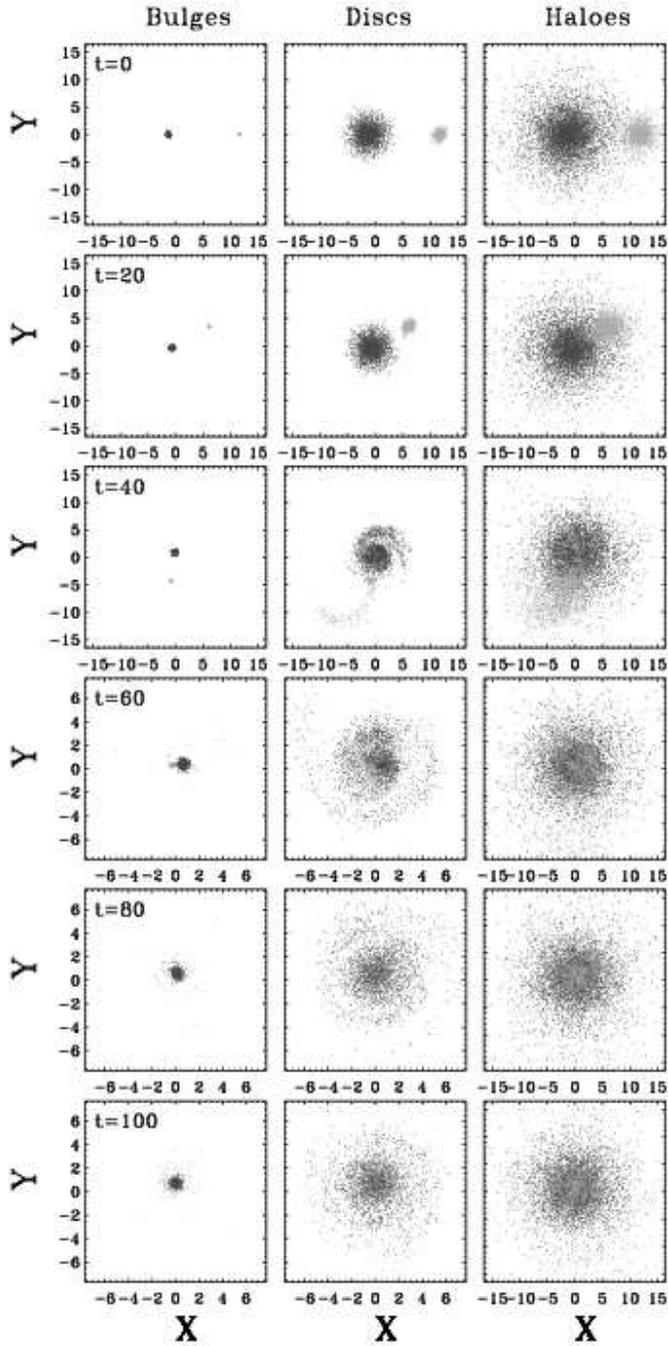}
\caption{Time evolution of the luminous material of model M3TF35D. Face-on views centered on the primary are plotted. Time is given in the upper left-hand corner of each frame. The total time of the simulation (100 time units) is equivalent to $\sim$2 Gyr, if the primary galaxy is scaled to match the Milky Way. First and second pericenter passages are at $t=28$ and $t=48$, respectively.  The $XY$ plane corresponds to the initial symmetry plane of the primary galaxy, except in the bottom snapshots where the galactic plane of the final remnant is plotted. Orbit is counterclockwise, prograde with the disc rotation. Only $1/15$ of the total number of particles is plotted in order to avoid saturation in the plot. {\it Black dots}: Primary particles.  {\it Gray dots}: Satellite particles. The bottom three snapshots for the luminous components have been zoomed-in.  
\label{Fig:faceon}}

\end{figure}

\begin{figure}[!tbp]
\centering
\includegraphics[width=\columnwidth]{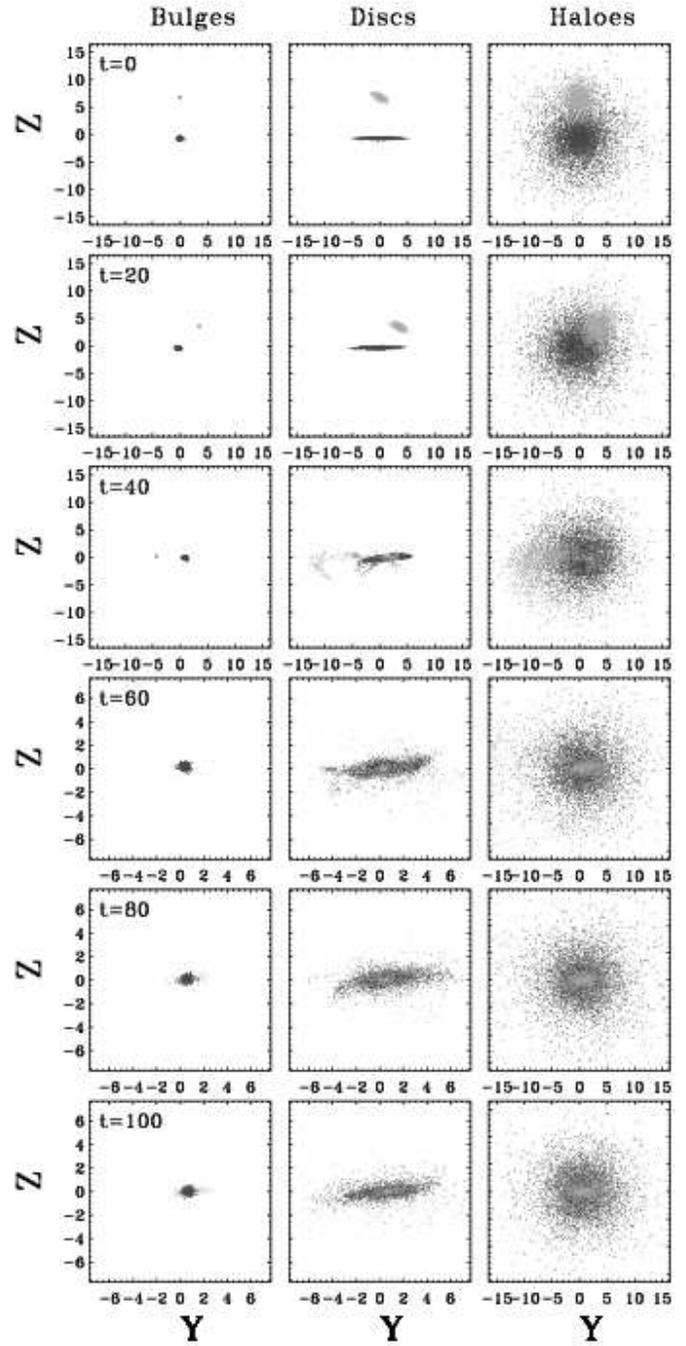}
\caption{Time evolution of the luminous material for model M3TF35D. Edge-on views centered on the primary are shown. Time is given in the upper left-hand corner of each frame.  The $Z$ direction is perpendicular to the initial galactic plane of the primary galaxy, except in the bottom snapshots where it is perpendicular to the galactic plane of the final remnant. Orbit is clockwise, prograde with the disc rotation. Only $1/15$ of the total number of particles is plotted in order to avoid saturation in the plot. See caption to Fig.\,\ref{Fig:faceon}. \label{Fig:edgeon}}
\end{figure}

\section{Results}

\label{Sec:results} 

The evolution of our systems is typical of unequal disc galaxy mergers, extensively described in the literature going back to \citet{Toomre72}. We show M3TF35D as an example in Figs.\,\ref{Fig:faceon} (face-on view) and \ref{Fig:edgeon} (edge-on view). Bulge, disc, and dark halo components are plotted separately for clarity.  Long tidal tails and a bridge connecting the two galaxies appear shortly after the initial pericenter passage at $t\sim 26$. 
In response to the tidal force from the companion, the inner regions of the primary disc develop transient spiral patterns and transitory non-axisymmetric distortions. At the end of the merger, satellite mass is distributed on a wide radial range over the primary disc.  We quantify the radial distribution in the following subsections.  Here we can already conclude that a  TF-scaling between primary and secondary can lead to significant disruption of the secondary.  Some material reaches the center of the remnant, and will contribute to the post-merger bulge light, but a large fraction ends up in the disc region. The contribution of satellites to the build-up of the thick disc had been shown in the low-density accretion experiment of ABP01, and also in the cosmological disc-galaxy evolution models of \citet{Abadi03a}.  

The evolution of retrograde mergers is similar to direct cases, except that tidal tails are inhibited due to the lack of spin-orbit coupling \citep{Mihos96}, and to the weak pericenter impulse, as pericenter separation is greater than the satellite tidal radius (GGB05).

\subsection{Surface density profiles}
\label{Sec:profiles}

\begin{figure*}[!]
\centering
\includegraphics[width=12cm]{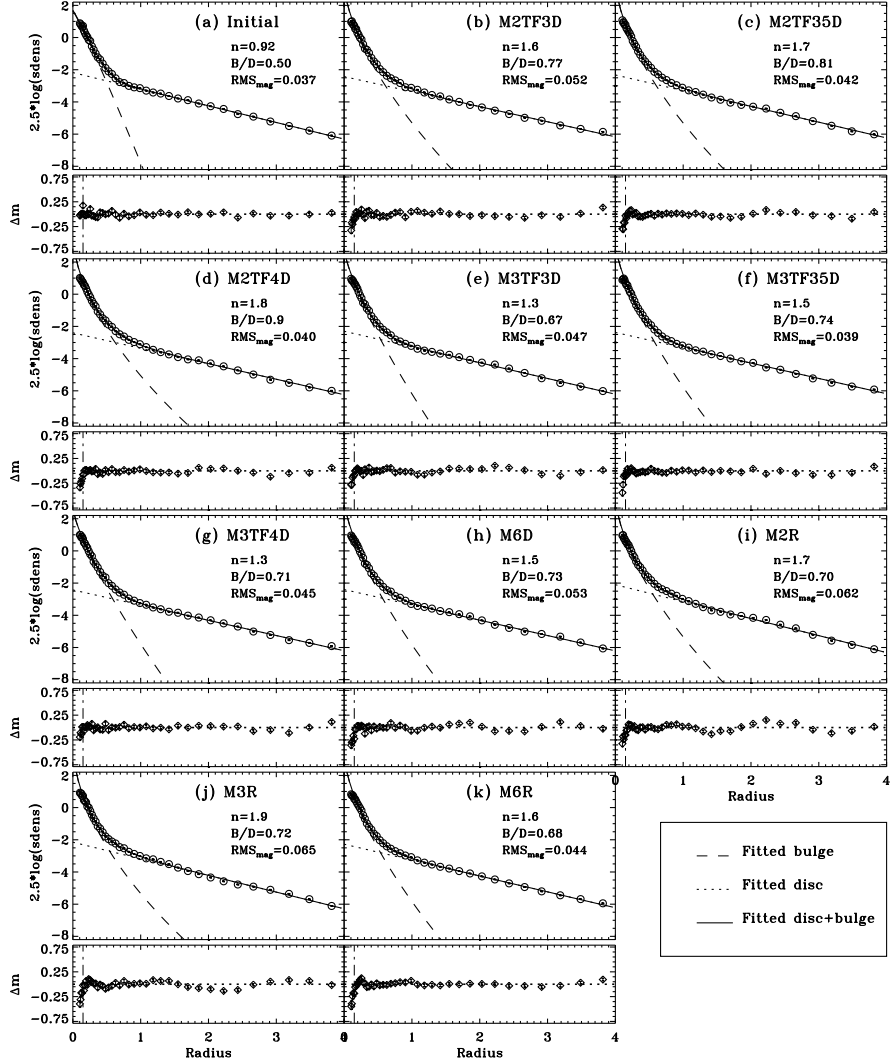}
\caption{Radial surface density profiles of luminous matter and S\'ersic+exponential simultaneous fits. \emph{Upper panels of each frame}: Surface density distributions and performed fits. Fitted values of the S\'ersic index $n$, $B/D$, and the RMS in magnitudes of the fit are shown in each panel. Panels: (\emph{a}) Initial primary model. (\emph{b-h}) Prograde models after the merger is complete. (\emph{i-k}) Retrograde models after the merger is complete. \emph{Open circles}: Model measurements with error bars. \emph{Dashed lines}: S\'ersic $r^{1/n}$ fitted component. \emph{Dotted lines}: Exponential fitted component. 
\emph{Solid lines}: Sum of the two fitted components. 
\emph{Lower panels of each frame}: Residuals profiles, in magnitudes. \emph{Diamonds}: Residuals of the fits in the above panel. Error bars of model measurements are also plotted on the residuals in magnitudes. The dashed vertical lines at $r = 0.15$ indicate the lower boundary of the fitted region.}  
 \label{Fig:sbr}
\end{figure*}

The accretion-driven evolution of bulge and disc structural parameters was inferred following steps similar to those used to infer bulge and disc parameters of real galaxies.  We derived azimuthally-averaged surface density profiles for the luminous matter, with all the merger remnants viewed face-on. 
We then fitted a combined S\'{e}rsic+exponential function to the density profiles. The S\'ersic law has been used extensively to model the surface brightness profiles of bulges \citep{Sersic68,Graham01a,Mollenhoff01,Prieto01,MacArthur02}.  Its functional form is 

\begin{equation}
I(r)=I_{\mathrm{e}}\cdot exp\,\{- b_{n}\, [ ( r/r_{\mathrm{e}}) ^{1/n}-1] \}  \label{Eq:Sersic}
\end{equation}
where $r_{\mathrm{e}}$ is the bulge effective radius, $I_{\mathrm{e}}$ is the surface density at $r_{\mathrm{e}}$, and $n$ is the S\'{e}rsic index. The factor $b_{n}$ is a function of the concentration parameter $n$, which, in the range $1<n<10$, may be approximated by $b_{n}=1.9992\, n-0.3271$, with an error $<0.15$\% in the whole range (see Fig.\,7 from G01). The exponential law provides a good model for discs \citep{Freeman70}. Its functional form is
\begin{equation}
I(r)=I_{\mathrm{0}}\cdot exp\left( -r/r_{\mathrm{0}}\right)  \label{Eq:disc}
\end{equation}
being $r_{\mathrm{0}}$ the scale length of the profile and $I_{\mathrm{0}}$ the extrapolated central surface density. The code used for fitting the radial surface density profiles was the one described by G01 with seeing-convolution turned off. The code uses a Levenberg-Marquardt nonlinear fitting algorithm to locate the $\chi ^{2}$ minimum by changing all model parameters. Points inside $r=0.15$ (the King core radius of the bulge) were excluded from the fit. We tested that including $\sim 5$ more points internal to the King core radius or excluding $\sim 5-10$ points external to it did not affect significantly the results of the fits \citep[see the study performed by][]{Aceves06}. The final face-on, elliptically-averaged radial surface density profiles of the luminous matter for all the models are shown in Figs.\,\ref{Fig:sbr}b-k. Figure~\ref{Fig:sbr}a is the surface density profile of the luminous matter for the primary galaxy before the accretion.

Residuals from the fits in magnitudes, defined as $\Delta m=2.5\log(sden/sdens_{fit})$, appear below each surface density profile plot in 
Fig.\,\ref{Fig:sbr}. Typical RMS is $\sim$0.05 mag, a 
quite reasonable result compared to typical observational errors. 
Secondary minima within the ocean of $\chi^{2}$ can arise from all the different model parameter combinations. We checked for their existence by changing the initial parameters by a factor of $\sim$10 typically.  We concluded that the $\chi ^{2}$ profiles had well-defined minima in all the cases. Final fitted parameters and bulge-to-disc mass ratios derived from the fits, with their errors \citep[bootstrap method: ][and references therein]{Efron93,Press94} are tabulated in Table~\ref{Tab:fits}. 

For all our experiments, we recover a two component system with the outer parts well fitted by an exponential profile (\ie, a disc), while the inner parts are well represented by S\'ersic fits with $n$ larger than 1.  The initial model presented $n\sim 0.9$. 

\begin{table*}
\begin{minipage}[t]{\textwidth}
\caption{Fitted parameters for S\'ersic bulge plus exponential disc decomposition of the
final merger remnants.}
\label{Tab:fits}

\centering
\renewcommand{\footnoterule}{}  
\begin{tabular}{cccccccccccc}
\hline\hline
       & & &\multicolumn{2}{c}{Disc}  & &
\multicolumn{3}{c}{Bulge}& &\\\cline{4-5}\cline{7-9}
Model Code& RMS &&log($\mu _{\mathrm{0}}$) & $h_{\mathrm{D}}$  & &log($\mu _{\mathrm{e}}$) &  $r_{\mathrm{e}}$  & {\it
n}  && $B/D$ \\
  (1)  & (2) &   & (3)  & (4)   & &(5)    & (6) &  (7)  &  &(8)\\\hline    
Initial&	0.037 &&  -0.871$\pm$0.016 & 1.051$\pm$0.023 &&   0.01$\pm$0.04 &   0.199$\pm$0.009 & 0.92$\pm$ 0.11 &&   0.50$\pm$0.03\\
M2TF3D& 	0.052 &&  -0.97$\pm$0.03   & 1.16$\pm$0.04   &&   0.11$\pm$0.06 &   0.192$\pm$0.013 & 1.62$\pm$ 0.30 &&   0.77$\pm$0.08\\
M2TF35D&	0.042 &&  -0.93$\pm$0.04   & 1.11$\pm$0.05   &&   0.16$\pm$0.07 &   0.182$\pm$0.014 & 1.71$\pm$ 0.36 &&   0.81$\pm$0.09\\
M2TF4D&		0.040 &&  -0.96$\pm$0.07   & 1.13$\pm$0.06   &&   0.12$\pm$0.13 &   0.191$\pm$0.023 & 1.78$\pm$ 0.80 &&   0.85$\pm$ 0.3\\
M3TF3D&		0.047 &&  -0.92$\pm$0.03   & 1.11$\pm$0.03   &&   0.13$\pm$0.08 &   0.184$\pm$0.015 & 1.32$\pm$ 0.34 &&   0.67$\pm$0.08\\
M3TF35D&	0.039 &&  -0.95$\pm$0.03   & 1.14$\pm$0.04   &&   0.16$\pm$0.07 &   0.182$\pm$0.012 & 1.47$\pm$ 0.29 &&   0.74$\pm$0.08\\
M3TF4D&		0.045 &&  -0.97$\pm$0.03   & 1.15$\pm$0.04   &&   0.07$\pm$0.06 &   0.200$\pm$0.012 & 1.32$\pm$ 0.23 &&   0.71$\pm$0.06\\
M6D&		0.053 &&  -0.96$\pm$0.04   & 1.14$\pm$0.05   &&   0.16$\pm$0.15 &   0.179$\pm$0.023 & 1.49$\pm$ 0.58 &&   0.73$\pm$0.15\\
M2R&		0.062 &&  -0.84$\pm$0.04   & 1.03$\pm$0.03   &&   0.11$\pm$0.07 &   0.187$\pm$0.014 & 1.68$\pm$ 0.51 &&   0.70$\pm$0.11\\
M3R&		0.065 &&  -0.86$\pm$0.05   & 1.05$\pm$0.05   &&   0.16$\pm$0.16 &   0.173$\pm$0.024 & 1.87$\pm$ 0.79 &&   0.72$\pm$0.18\\
M6R&		0.044 &&  -0.91$\pm$0.03   & 1.10$\pm$0.03   &&   0.15$\pm$0.08 &   0.175$\pm$0.015 & 1.56$\pm$ 0.32 &&   0.68$\pm$0.08\\[0.05cm]\hline\\
\end{tabular}  
\begin{minipage}[t]{\textwidth}
\emph{Description of columns}:  
(1) Model code. 
(2) RMS of the fit in magnitudes. (3) Disc central surface density. (4) Disc scale length. (5) Bulge effective surface density. (6) Bulge effective radius. (7) Bulge profile S\'ersic index. (8) Bulge-to-disc mass ratio derived from the S\'ersic+exponential
fit.
\end{minipage}
\end{minipage}
\end{table*}

\begin{figure*}[!]
\centering
\includegraphics[width=12cm]{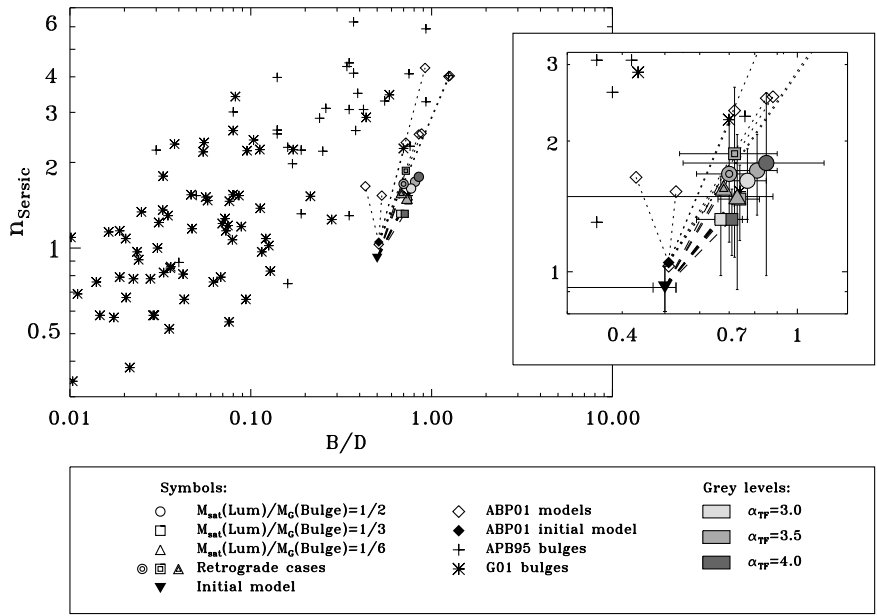}
\caption{Growth vectors in the log\,($n$)-log\,($B/D$) plane for the present TF-scaled models. Right panel shows a zoom of the left panel. Each segment starts at the location of the original model and ends at the $n$ and $B/D$ derived from the two-component fit to the surface density profile after the merger. Errors from Table~\ref{Tab:fits} are overplotted. Growth vectors of ABP01 high-density models are plotted for comparison. Distributions of $\log\,(n)$ vs.\,log($B/D$) for bulges of several studies are drawn too. Symbols and colours are explained in the legend. 
\label{Fig:growth}}
\end{figure*}

\subsection{Growth of bulges}
\label{Sec:growthofbulges}

Figure~\ref{Fig:growth} shows bulge \emph{growth vectors} in the plane log\,($n$) vs.\,log\,($B/D$). Each segment originates at the location of the pre-merger primary bulge, and ends at the location of the post-merger bulge in the $\log\,(n)$ vs.\,$\log\,(B/D)$ plane.  
$B/D$ and $n$ derive from the S\'{e}rsic+exponential fits (Table~\ref{Tab:fits}).  
We also show growth vectors for the dense satellites accretion models of ABP01, as well as real bulges from the samples of \citet{deJong96b}, re-analysed by G01, and APB95. 

In all our models, \emph{satellite accretion leads to a simultaneous increase of $B/D$ and of the S\'ersic index of the bulge.}  ABP01 had found a similar trend for galaxies ingesting a dense, spheroidal satellite.  The present models show that satellites with disc structure and TF-scaled densities also  lead to a correlated increase in $n$ and $B/D$.  We will show in \S~\ref{Sec:whynincreases} that the processes leading to the increase of $n$ and $B/D$ are different for the two sets of models. Here, we compare the two trends.  ABP01 and our experiments show similar features: we find a tendency for more massive satellites to lead to higher $B/D$ and $n$; and, the satellite density, parametrized by \alphaTF, has little effect on the bulge growth for the range of TF exponents studied here. Quantitatively however, there are differences. The present models have lower $n$ indices for the same final $B/D$. The trends of ABP01 models and ours seem to lie along  sequences given by  $n = \beta\,(B/D)^{\,\alpha}$. As we wanted to force the line  $\log\,(n)$-$\log\,(B/D)$ to pass through the point corresponding to the initial model, we have used a gradient-expansion algorithm to compute a non-linear least squares fit to the present models using the following expression:
\begin{equation}
\log\,\left[\frac{n}{n_\mathrm{\,ini}}\right] = \alpha\,\log\,\left[\frac{(B/D)}{(B/D)_\mathrm{\,ini}}\right]
\label{eqn:NvsBD}
\end{equation}
where $(B/D)_\mathrm{\,ini}$ and $n_\mathrm{\,ini}$ represent the values corresponding to the initial primary galaxy (see Table~\ref{Tab:fits}). Our TF-scaled models fitted eq.\,\ref{eqn:NvsBD} for $\alpha = 1.4\pm 0.8$, with $\chi ^{2}=0.021$. So, roughly, in the present TF-scaled models $n$ increases by a factor of $\sim 2.6$ when $B/D$ doubles from $n=0.9$ (although our experiments do not reach such a high increment in $B/D$, see Table~\ref{Tab:fits}). However, the ABP01 remnants obeyed worstly the trend given by eq.\,\ref{eqn:NvsBD}, exhibiting $\chi ^{2}=0.090$ for $\alpha = 1.7\pm 0.7$. This means that, in ABP01 models, $n$ increased by a factor of $\sim 3.2$ (from $n=1$) when $B/D$ doubled.  The gentler increase in $n$ found in the present experiments is welcome, as it alleviates the paradox posed by the suggestion from ABP01 models that exponential bulges are fragile to merging, given that  galaxies with exponential bulges are very common.  In the light of the present models, we see that the fragility of exponential bulges stemmed from the use of very dense satellites; satellites with lower densities, as in our TF-scaled models, lead to a gentler $n$ increase as $B/D$ grows.  The maximum $n$ in our set of models is $n=1.87$, whereas ABP01 high-density experiments got up to $n=4.30$.

Observed bulges are distributed along a sequence of increasing $n$ and $B/D$ (Fig.~\ref{Fig:growth}).  Our results indicate that galaxies in the lower-left of the diagram,  (low-mass, $n\approx 1$ bulges), would move toward the upper-right (massive, high-$n$ bulges) when accreting a low-mass galaxy.  ABP01 ran an experiment that showed the cumulative effects of merging (see their model H3D3D). They found exactly the same increase in $n$ and in $B/D$ in two consecutive mergers with exactly the same mass (1:3) onto their primary galaxy. Therefore, if the effects of merging are cumulative, as this experiment seems to indicate, satellite accretion provides an evolutionary link connecting low-$n$, small bulges to more massive, high-$n$ bulges.  Moreover, \citet{Khochfar05} found that minor  mergers are a magnitude more frequent than major mergers, a result that supports the possible relevance of minor mergers in the configuration of actual bulges. 


We caution that our models start out with higher $B/D$ than most real galaxies (Fig.~\ref{Fig:growth}).  This offset mainly derives  from computational limitations: the higher range of dynamical times in galaxies with smaller bulges lead to more time-consuming computations; and, accretion experiments onto smaller bulges require smaller satellites, which take longer to decay. While we expect the main trend of increasing $n$ and $B/D$ to remain valid in the low-$B/D$ regime, the details will probably change.  Small, $n < 1$ bulges should have a weaker tidal field which probably allows for smaller satellites to reach the center undisturbed.  This might lead to a steeper $n$-$\log(B/D)$ trend than the one derived from the present models  (eq.~\ref{eqn:NvsBD}).  Conversely, we expect shallower slopes for the $n$-$\log(B/D)$ trend for more massive bulges that start off with $n>1$; the process of $n$ growth via dissipationless accretion saturates at  some value of $n$, probably around $n\sim 4$ (ABP01).  Shallower growth vectors for more massive bulges would keep accretion-grown bulges within the locus of real bulges in the $n$-$\log(B/D)$ diagram.

\begin{figure*}[!]
\centering
\includegraphics[width=12cm]{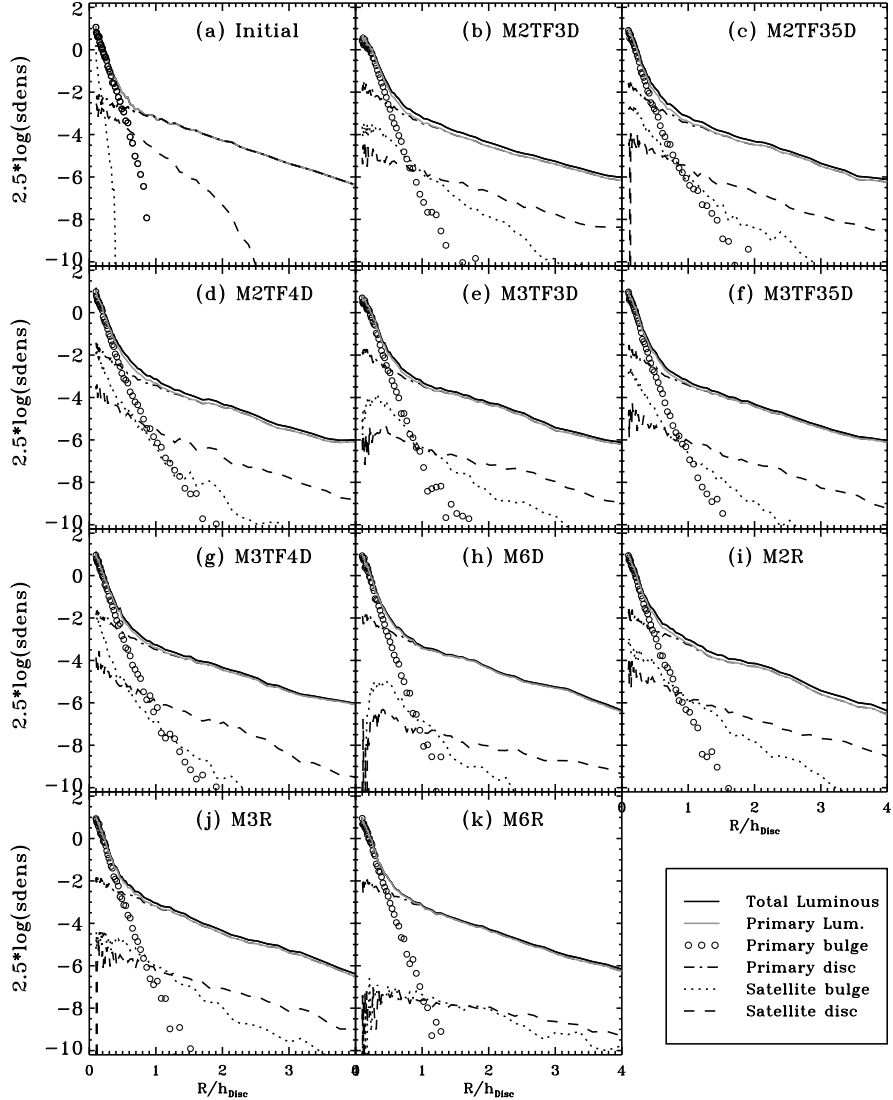}
\caption{Face-on radial surface density distributions of the various luminous components, for all the models after the merger. \emph{Solid lines}: Total luminous material. For the initial model, only luminous material from the primary galaxy is considered. \emph{Open circles}: Primary bulge material. \emph{Dashed-dotted line}: Primary disc material.  \emph{Dotted line}: Satellite bulge material. \emph{Dashed line}: Satellite disc material. \label{Fig:sbrcomponents}}
\end{figure*}

\subsection{Why does $n$ increase}
\label{Sec:whynincreases}

We now analyze the radial mass distributions per component for the models, in order to clarify what drives the increase of $n$ and $B/D$.  In Fig.\,\ref{Fig:sbrcomponents} panels (b)-(k), we plot the final surface density profiles for particles originally belonging to  each mass component, for all of our remnants; panel (a) corresponds to the primary and the satellite scaled with $\alphaTF =3.5$ models before merging.  Three distinct processes modify the mass distribution of the galaxies.   First, satellite material gets deposited over a wide radial range in the remnants.  Second, the distribution of primary bulge material becomes more extended in its outer parts.  And, third, the primary disc material becomes more centrally concentrated:  in all of the models, surface density profiles are not exponential but curve up inwards.  We show here that primary disc redistribution is the dominant contributor to the change in the galaxy profile.  

We already noted earlier in \S\,\ref{Sec:results} that satellite material gets deposited over a wide radial range.  Figure~\ref{Fig:sbrcomponents} shows that, in the $\alphaTF=3.5,4.0$ cases, satellite material does reach the center of the remnant, while low-density ($\alphaTF=3.0$) or low-mass (M6) satellites entirely disrupt during the merger; their mass gets deposited into an extended, torus-like structure. But the contribution of this material to the total profile is small. In no case does the satellite material dominate the central surface density. Rather, its contribution to the total surface density is between $\sim 2-6$ mag below that of the primary. Only for  $\alphaTF=4.0$, the $r=0$ contribution of the satellite bulge is similar to that of the primary disc.  To highlight the small contribution of accreted material, the surface density profile from primary material alone is also shown in Figure~\ref{Fig:sbrcomponents} (solid gray lines), together with the total luminous profile (solid black lines).  The difference between both profiles, which measures the contribution of the satellite material, never reaches 0.2 mag\,arcsec$^{-2}$.  Also, excluding the secondary material modifies the parameters from the S\'ersic+exponential fits by less than $\sim 20$\%.   The situation is, therefore, quite unlike that found for high-density, spheroidal satellites by ABP01, in which the central density was dominated by accreted material.  

\begin{figure*}
\centering
\includegraphics[width=12cm]{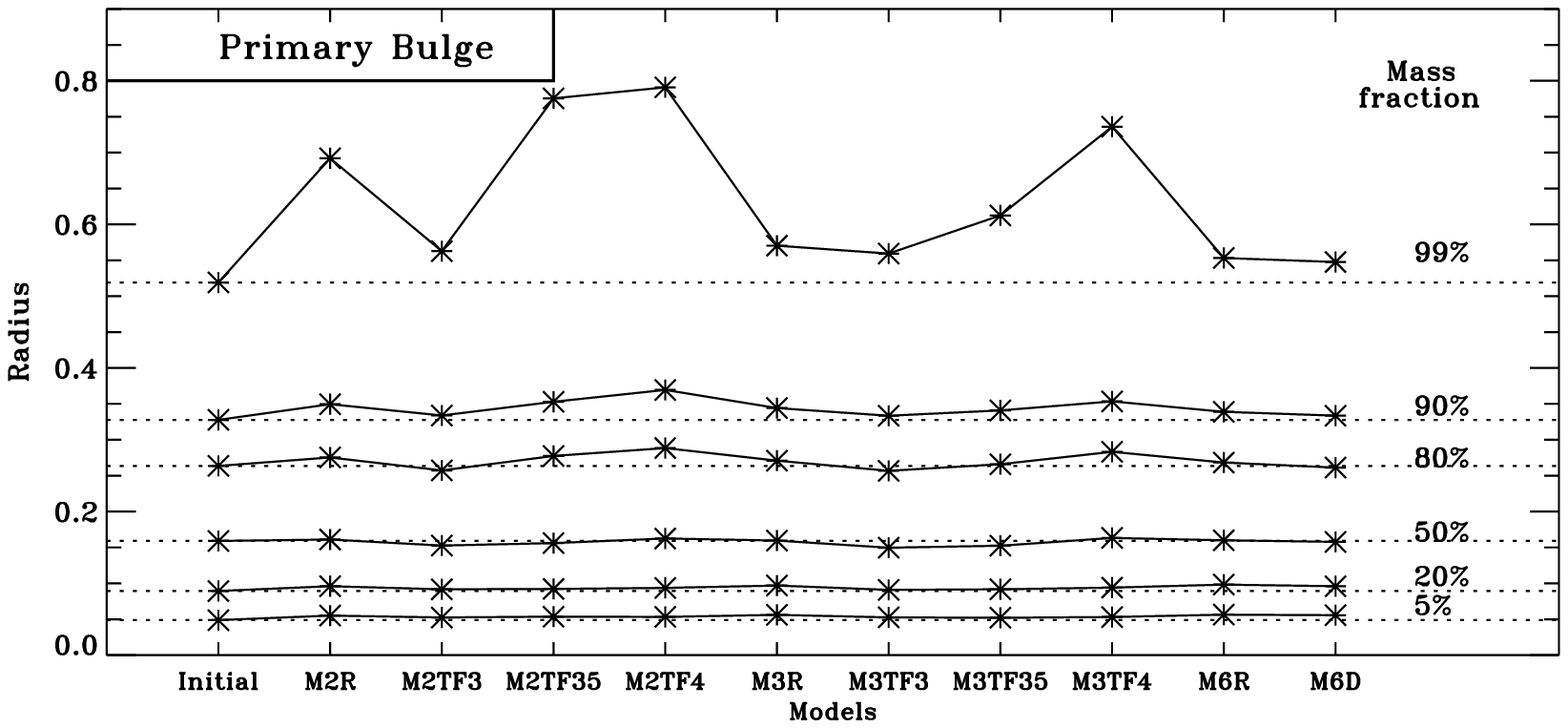}
\includegraphics[width=12cm]{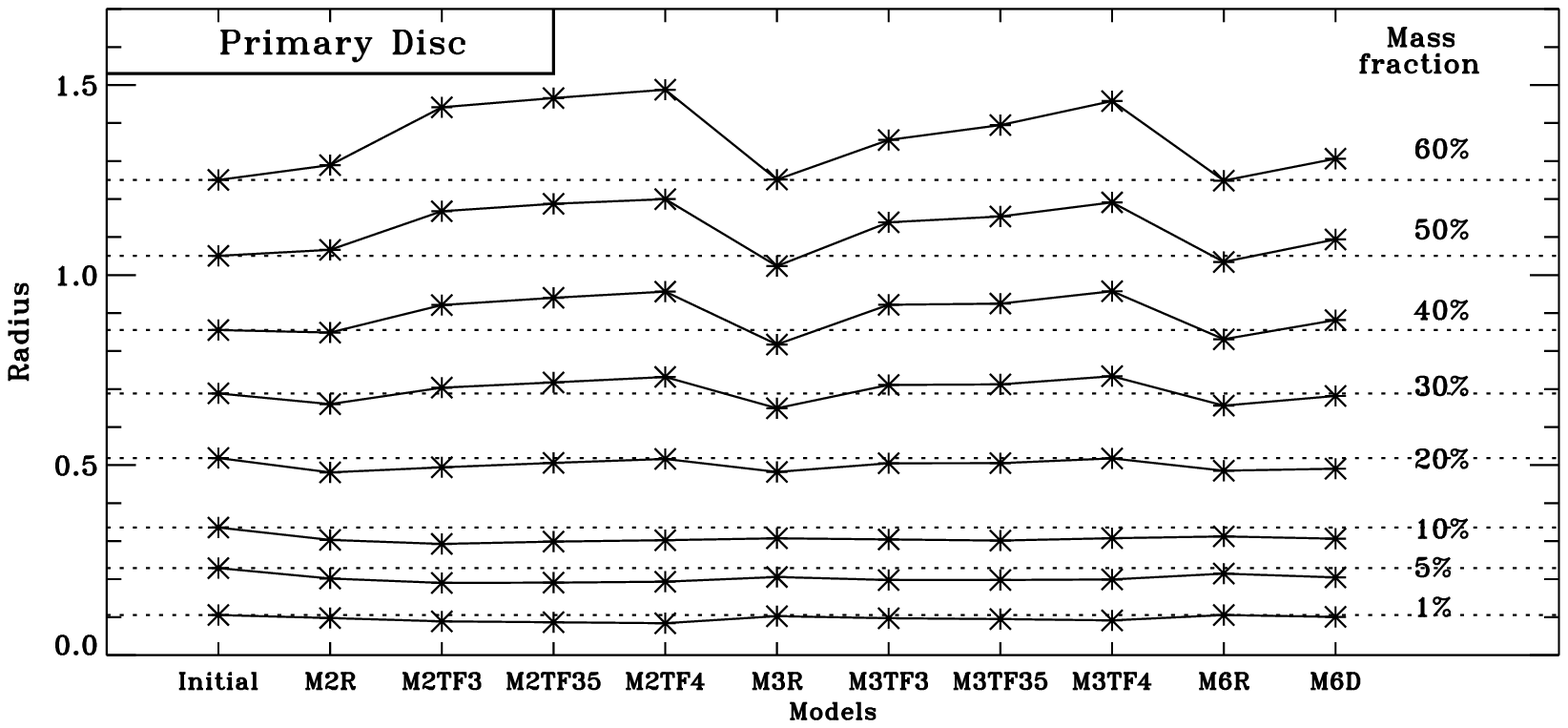}
\caption{Radii enclosing a given \% of the mass of the primary galaxy for all the models.  The abcissa values are the model codes from Table~\ref{Tab:orbits}. The horizontal dotted lines are the radii initially enclosing a given \% of the mass, which appears next to the line, on the right of the figure. \emph{Upper panel}: For the particles initially belonging to the primary bulge. \emph{Lower panel}: For the particles initially belonging to the primary disc. \label{Fig:bulge}}
\end{figure*}

To investigate which of the two primary-galaxy luminous components has the dominant contribution to the surface density profile change, we analyze fractional mass distributions (FMD), i.e., the radii enclosing given percentages of the mass.  Figure~\ref{Fig:bulge} shows the FMD for the primary bulge particles (top panel) and for the primary disc particles (lower panel), for all of the models.  For the primary bulge, expansion occurs in the outer 20\% of the mass, at radii from 0.2 to $\sim$1.0, and is strongest for high-density (TF4) satellites; the inner 70\% of the mass remains almost undisturbed in all cases. For the primary disc, FMD changes at all radii.  The inner 20\% contract (radii $r \leq 0.8$) while the outer mass fractions tend to expand.  

Hence, \textit{the changes in the final surface brightness profiles are mostly driven by the redistribution of the disc material}.  By increasing its central concentration, the disc adds material to the bulge region and increases its central surface density; and, by becoming brighter at radii where the profiles change from being bulge-dominated to disc-dominated, the disc material gives the final profile a smooth curvature that leads to higher-$n$ S\'ersic fits.  The changes in the primary bulge FMD have a secondary effect on the final profiles, as expansion occurs at radii where their contribution to the total profile is several magnitudes below that of the primary disc. The physical process driving the expansion of the primary bulge outer regions is the absorption by the bulge material of the orbital energy and angular momentum. Bulge expansion is greatest for the denser (TF4) satellites (Fig.~\ref{Fig:bulge}); but, in general, it is lower here than in the ABP01 dense spheroidal models, due to the lower mass, lower density and progressive disruption of the satellites we have modelled.  For the disc, implosion of the inner 20\% of its mass is due to the effects of the transitory non-axisymmetric distortions driven by the tidal field of the satellite, as already described by \citet{Mihos94}, and to the deposition of dark matter from the satellite halo in the center of the remnant. Notice also that the primary disc outer parts expand, contributing to the increase of the disc scale lengths in the final remnants (see lines for FMD $\ge 30$\% in the bottom panel of Fig.\,\ref{Fig:bulge}).

\subsection{Population mixing}
\label{Sec:populationmixing}
 
We now quantify the amount of population mixing in the region of the bulge induced by the satellite infall.  For the purpose of this analysis we simply define the region of the bulge as $r\leq 0.5$ (cf.~Fig.~\ref{Fig:sbr}), and study the evolution of the parameter

\begin{equation}
\eta_{i}(t) \equiv \frac{M_{i}(t)}{M_{\mathrm Lum}(t=0)}
\label{eqn:contrib2bulge}
\end{equation}
where $M$ denotes mass within $r=0.5$, $t$ is time, and $i=$ B1,D1,B2,D2 for primary bulge, primary disc, secondary bulge and secondary disc material, respectively. We have called $M_{\mathrm Lum}(t=0)$ the total primary luminous mass that initially existed at $r\leq 0.5$. Therefore, $\eta_\mathrm{total}(0)=1$, but $\eta_\mathrm{total}(t_\mathrm{final})\ne 1$. The disc contribution to the region of the bulge, initially $\etaDpri(0) = 0.16$, rises in all of the models, in proportion to the satellite mass: $\etaDpri(t_\mathrm{final}) =$ 0.18, 0.19, and 0.20 for mass ratios 1:6, 1:3, and 1:2, respectively.  These numbers are slightly lower for retrograde orbits, and do not depend on the TF exponent.  
The bulge contribution, initially $\etaBpri(0) = 0.84$, decreases in all cases in proportion to the satellite mass and also to the TF exponent, down to $\etaBpri(t_\mathrm{final}) \sim 0.79$ for the most massive and densest case (M2TF4D).  Finally, the satellite bulge particles accumulate in the bulge region, whereas the satellite disc material is deposited mostly in regions  external to the bulge.  As expected, deposition in the bulge region is stronger for more massive or denser satellites, and for direct orbits.   Highest values (M2TF4D) are $\etaBsec (t_\mathrm{final})= 0.09$ and $\etaDsec (t_\mathrm{final}) = 0.03$.  

\begin{figure*}[!]
\centering
\includegraphics[width=12cm]{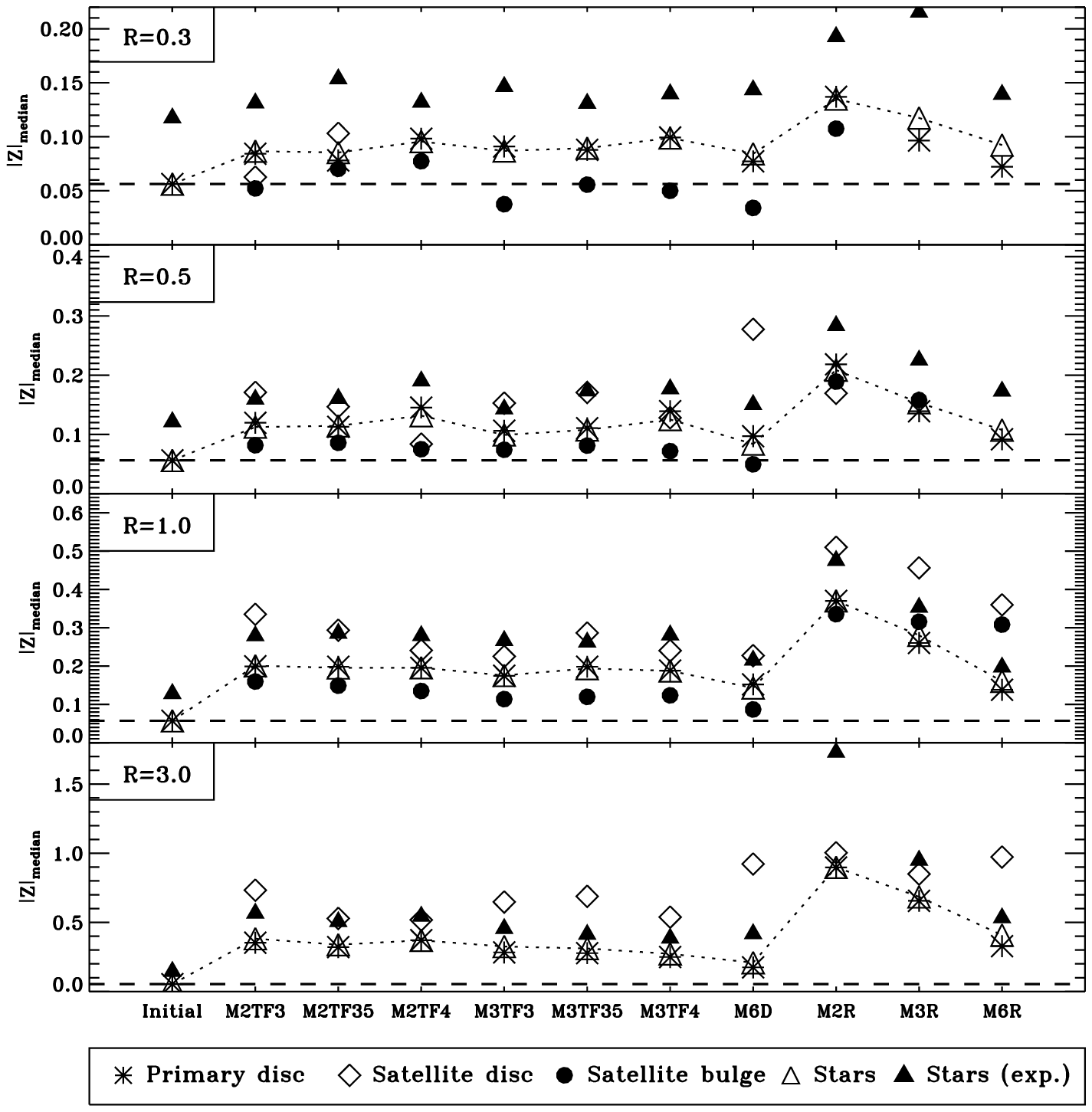}
\caption{Medians of the vertical distributions in the final remnants for the particles initially belonging to different components, at four fixed radii in the galactic plane ($r=0.3, 0.5, 1.0, 3.0$). The abcissa values are the model codes from Table~\ref{Tab:orbits}. The horizontal  dashed lines in each panel indicate the initial vertical scale of the primary disc at each radius. Only data points with more than 50 particles per bin are plotted. \emph{Asterisks}: Vertical scales for the particles initially belonging to the primary disc. \emph{Diamonds}: Vertical scales for the particles initially belonging to the satellite disc. \emph{Filled circles}: Vertical scales for the particles initially belonging to the satellite bulge. \emph{Open triangles}: Vertical scales for all the luminous material. \emph{Filled triangles:} Scale heights obtained through exponential fits to the vertical mass distribution of the total luminous matter in each remnant, at the fixed radius.\label{Fig:zscale}}
\end{figure*}

\subsection{Vertical structure of the remnants}
\label{Sec:thickening} 

Simulations have shown that merging satellites with masses as low as a few percent of the disc mass heat discs efficiently \citep{Quinn93,Mihos94,Huang97}.  As pointed out by \citet{Toth92}, disc thickening effects put constraints on how important merging processes may have been in the history of disc galaxies.  Our models are not an exception: the final disc of all the remnants exhibits higher vertical scale than the initial disc. In this subsection we provide measurements of disc thickening for our models.  These measurements serve two purposes.  First, extreme thickening would falsify the hypothesis that satellite merging is relevant for the evolution of bulges of disc galaxies.  And, second, our analysis will highlight that accreted satellites are important contributors to the building of thick discs \citep{Abadi03b}.

In Figure\,\ref{Fig:zscale} we plot the medians of the final vertical distributions ($\left| z\right| _\mathrm{median}$) for all the models, computed at four fixed radial positions ($r=0.3, 0.5, 1.0$, and 3.0), and separately for particles initially belonging to the primary disc, the satellite bulge, and the satellite disc. We have also plotted the scale heights obtained through exponential fits to the vertical distribution of the total luminous material at each fixed radius. We see that scale heights increase for all the models and all the mass components.  Focussing on the $r=1$ panel, for the primary disc particles,  $\left| z\right|_\mathrm{median}$ increases by factors of 2 to 4 from an initial value $\left| z(t=0)\right|_\mathrm{median}\sim 0.05$.  


For the satellite particles, Fig.\,\ref{Fig:zscale} shows that $\left| z\right| _\mathrm{median}$ at $r \gtrsim 0.5$ is higher that $\left| z\right| _\mathrm{median}$ for primary disc particles.  I.e., the material accreted from the satellite disc contributes to building a thick disc.  In contrast, the satellite bulge particles are confined for all the remnants in a central ($R<2$), thin (low $\left| z\right| _\mathrm{median}$) distribution. We will show in \S\ref{Sec:dynamics} that such structures are dynamically-cold inner discs. 

Figure~\ref{Fig:zscale} also shows that, for particles from the primary and secondary discs, scale heights increase with radius (note that the ranges plotted in the vertical axes change from panel to panel). I.e., our final discs are flared. Disc flares and warps are common in real discs \citep{deGrijs97}, and can be easily reproduced in simulations of merging small satellites \citep{Quinn93}.

The values plotted in Fig.\,\ref{Fig:zscale}  are raw measurements, which need to be corrected for numerical thickening which arises from the residual collisionality of the code.  To estimate this correction, we assume that energy injection into vertical motions in the disc, coming from two-body effects and from merger dynamics, are largely independent from each other:   two-body disc vertical heating is dominated by interactions with primary halo particles, whose effects should not vary much if the disc is been thickened by the satellite accretion; and, accretion-driven injection of vertical kinetic energy onto the disc should not vary much under small variations of the disc thickness due to halo-disc two-body effects.  Under that assumption, we have 

\begin{equation}\label{eq:kineticenergy}
K_{z,\mathrm{simul}} = K_{z,\mathrm{initial}} + \Delta K_{z,\mathrm{2B}}+\Delta K_{z,\mathrm{merging}}
\end{equation}
or
\begin{equation}\label{eq:dispersion}
\sigma^2 _{z,\mathrm{simul}} = \sigma^2 _{z,\mathrm{initial}} + \Delta \sigma^2 _{z,\mathrm{2B}}+\Delta \sigma^2 _{z,\mathrm{merging}}
\end{equation}
where $K_{z}$ denotes vertical kinetic energy, $\sigma_{z}$ denotes vertical velocity dispersion, and \textit{initial, 2B, merging, simul} denote initial, two-body, merger, and final system, respectively.  Because vertically the discs are approximately isothermal, $h_{z} = \sigma_{z} / (2\,\pi\,G\,\rho_{0})^{-1/2}$ \citep{Spitzer42,vdKruit81}, where $\rho_{0}$ is the in-plane volume density.  In the limit of low mass accretion valid in the present experiments, vertical disc surface density is conserved, i.e., $\mu \propto \rho_{0}h_{z} = \mathrm{const.}$, hence  
$h_{z} \propto \sigma_{z}^{2}$, and equation~\ref{eq:dispersion} becomes
$h _{z,\mathrm{simul}} = h _{z,\mathrm{initial}} + \Delta h _{z,\mathrm{2B}}+\Delta h _{z,\mathrm{merging}}$. Under these approximations thickening effects add up linearly, i.e., 

\begin{equation}\label{eq:scaleheight}
\Delta h _{z,\mathrm{merging}} \approx \Delta h _{z,\mathrm{simul}} - \Delta h _{z,\mathrm{2B}}
\end{equation}
\noindent where $\Delta h _{z,\mathrm{simul}}$ is given by Fig.\,\ref{Fig:zscale}.
We measured $\Delta h _{z,\mathrm{2B}}$ by evolving the initial primary galaxy model in isolation for 100 time units, the typical duration of the experiments.  We obtained 
$\Delta h _{z,\mathrm{2B}} = 0.05$, a similar result to \citet{Kuijken95}.  From inspection of Fig.~\ref{Fig:zscale}, we conclude that $\Delta h _{z,\mathrm{merging}} \approx 0$ for the low-mass satellites, where the measured thickening is dominated by two-body effects.  For most models, $\Delta h _{z,\mathrm{merging}} / h_{z,\mathrm{initial}} \approx 3$, and $\Delta h _{z,\mathrm{merging}} / h_{z,\mathrm{initial}} \approx 6$ for the most massive, retrograde experiments.   

Clearly the post-merger systems do not match thin discs of present-day spirals.  Their properties resemble those of thick discs.  Assuming the rebuilding of thin discs out of left-over gas after the merger, disc thickening is not a problem for the merger origin of bulge growth.  The build-up of a thick disc may be a useful feature of our models for evolution toward earlier Hubble types, as thick discs appear to occur more often in early types than in later types \citep{Burstein79}.  Finally, our results agree with \citet{Abadi03b} that accreted satellites contribute to building the thick disc; we note, however, that the set of orbit inclinations covered by our experiments is rather limited. 

\begin{figure*}[!]
\centering
\includegraphics[width=12cm]{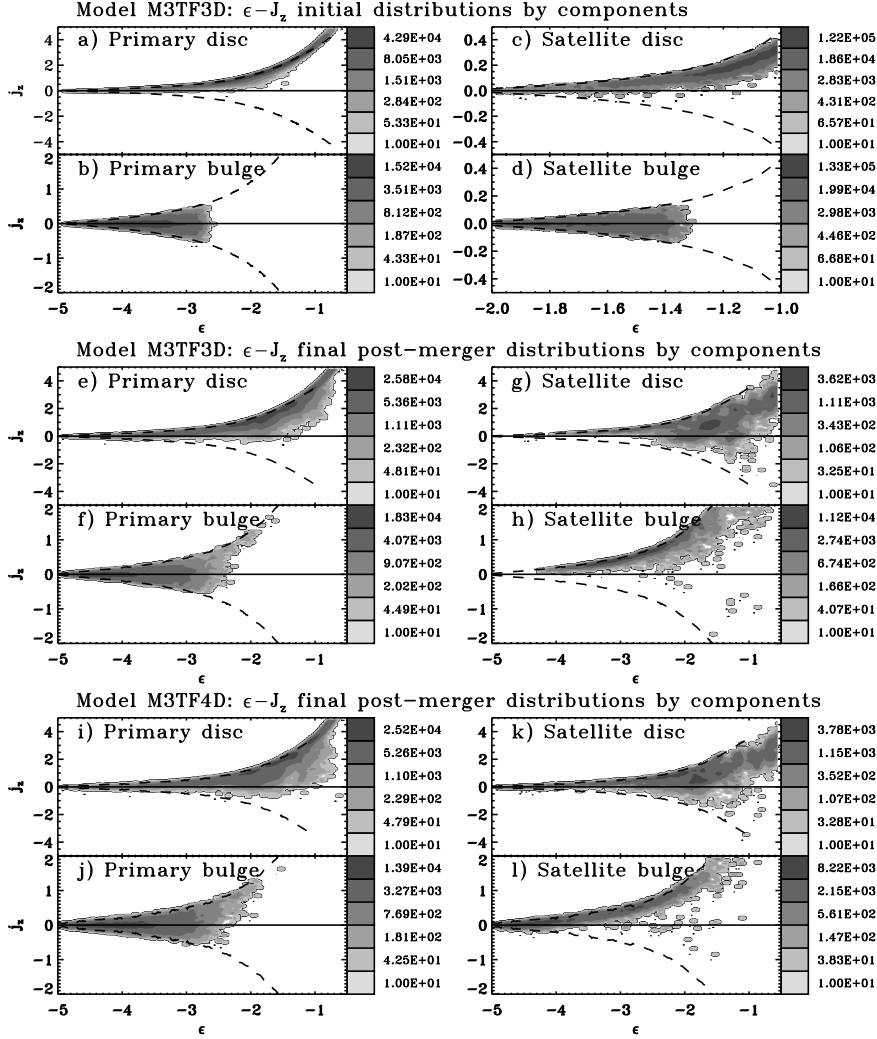}
\caption{Initial and final particle density distributions in the plane $j_{\mathrm{z}}$-$\varepsilon$ for each galactic component of models M3TF3D and M3TF4D, being $j_{\mathrm{z}}$ the $z$-component of the specific angular momentum, and $\varepsilon$, the specific energy. Darker grey levels in the density contours point to higher number of particles populating the corresponding  $j_{\mathrm{z}}$-$\varepsilon$ bin (see the corresponding legend on the right of each panel).  Contours are logarithmically-spaced. \emph{Panels (a-d)}: Initial material distributions in the $\varepsilon -j_\mathrm{z}$ plane of the two luminous components of the primary galaxy and the satellite for model M3TF3D (initial distributions for all the models are analogous). \emph{Panels (e-h)}: Post-merger distributions for the two luminous components of the primary galaxy and the satellite in the remnant of model M3TF3D. \emph{Panels (i-l)}: Post-merger distributions for the two luminous components of the primary galaxy and the satellite in the remnant of model M3TF4D. \emph{Dashed lines}: Locus of circular orbits co-rotating ($j_{\mathrm{z}} >$0) and counter-rotating ($j_{\mathrm{z}} <$0), $j_{\mathrm{circ}}(\varepsilon )$, in each case. \label{Fig:energyjz}}
\end{figure*}

\begin{figure*}[!]
\centering
\includegraphics[width=12cm]{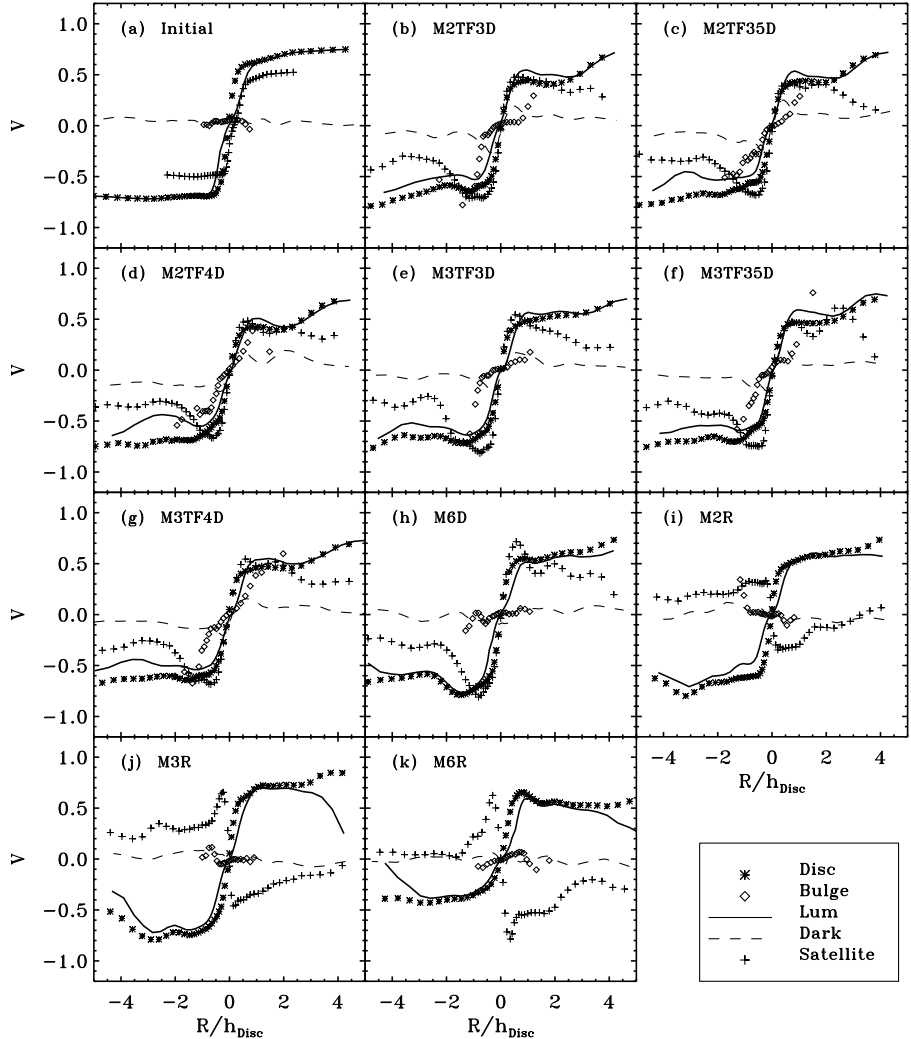}
\caption{Line-of-sight velocities of the models, using a slit placed onto the disc major axis of the remnant, for the inclined view: $\theta = 60^\circ$, $\phi =90 ^\circ$. The abcisas axis has been inverted in the retrograde models. The curve labelled as "satellite" comprises all the luminous material initially belonging to the satellite. (\emph{a}) Primary and secondary initial models. (\emph{b-h}) Prograde models after the merger is complete. (\emph{i-k}) Retrograde models after the merger is complete. \emph{Asterisks}: Primary disc material. \emph{Diamonds}: Primary bulge material. \emph{Crosses}: Satellite luminous material (bulge+disc). \emph{Solid line}: Total luminous material (satellite+primary galaxy), except for panel (a), where only total luminosity material from the primary galaxy is plotted. \emph{Dashed line}: Total dark matter halo (satellite+primary galaxy), except for panel (a), where only dark matter from the primary galaxy is plotted.\label{Fig:rotcurve1}}
\end{figure*}

\begin{figure*}[!]
\centering
\includegraphics[width=12cm]{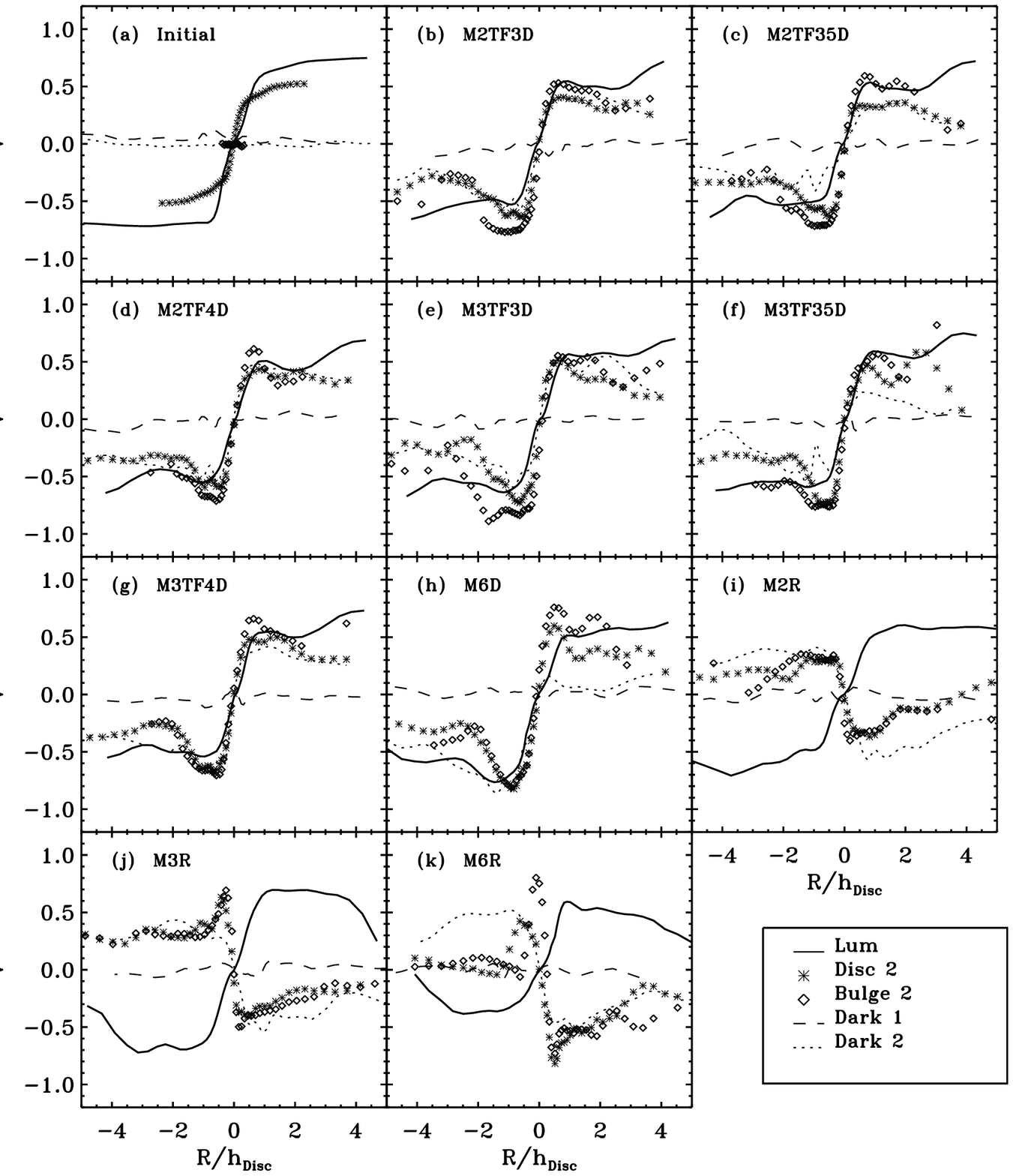}
\caption{Line-of-sight velocities of the particles originally in the satellite, after the merger is complete, for the inclined view: $\theta = 60^\circ$, $\phi =90 ^\circ$. \emph{Asterisks}: Satellite disc material. \emph{Diamonds}: Satellite bulge material. \emph{Solid line}: Total luminous material (satellite+primary galaxy), except for panel (a), where only total luminosity material from the primary galaxy is plotted. \emph{Dashed line}: Primary galaxy dark material. \emph{Dotted line}: Satellite dark material. \label{Fig:rotcurve2}}
\end{figure*}

\subsection{Phase-space evolution}
\label{Sec:dynamics} 

The structural transformations discussed in \S~\ref{Sec:whynincreases} must be the result of changes in phase-space structure of the galaxies.  We show that this is the case by looking at Lindblad (energy-angular momentum) projections of phase space. Figure \ref{Fig:energyjz} presents specific angular momentum vs.\ energy diagrams ($j_{\mathrm{z}}$ - $\varepsilon$) before and after the accretion, for each galaxian component of models M3TF3D and M3TF4D. These two models are useful for showing the effects of satellite density (higher for $\alphaTF=4.0$) on the final dynamical structure of the remnants.  Top panels of the figure show the initial particle distributions of the luminous components of the primary galaxy and the satellite. The four intermediate panels present the final post-merger particle distributions of model M3TF3D, and the bottom panels, the same but for M3TF4D. Particle densities in the Lindblad diagrams are depicted with logarithmically-spaced levels to highlight low-level structures.  The $z$ direction is defined by the total spin of all the luminous material of the primary galaxy (panels [a]-[b]); the satellite (panels [c]-[d]); and the final remnant (panels [e]-[h] and [i]-[l]). The thick dashed line in each panel indicates the location of circular orbits,  $j_{\mathrm{circ}}(\varepsilon)$, corresponding to the potential of the primary galaxy (panels [a]-[b]), the satellite (panels [c]-[d]), or the final remnant of each model (panels [e]-[h] and [i]-[l]). 

Figure \ref{Fig:energyjz} shows that material from the primary galaxy modifies its distribution onto the $\varepsilon -j_\mathrm{z}$ plane only moderately. Disc particles, which were initially distributed in nearly circular orbits, scatter over the region of direct orbits allowed by the potential (compare panel [a] with panels [e] and [i] in Fig.\,\ref{Fig:energyjz}). Bulge particles, with no initial rotation,  acquire rotation in their outer regions ($\varepsilon \gtrsim -3.0$; see panels [b] and [f],[j]).  Both processes bring the two distributions to look more like each other, i.e., the accretion leads to a mixing of bulge and disc in phase space.  The changes in surface brightness profile and in the vertical structure of the galaxy (\S\S\,\ref{Sec:growthofbulges} and \ref{Sec:populationmixing}) are consequences of such mixing in phase space.  The transformation is gentle, as the global energy and angular momentum ranges do not change during the accretion.  

Phase-space evolution is much stronger for particles from the satellite.  Material from the satellite disc, originally lying along the locus of circular orbits, end up nearly filling the available prograde phase space, and populating some retrograde orbits (compare panel [c] with panels [g] and [k]).  Thus, the secondary disc is destroyed.  That orbits in the final potential are not circular is a result of an early disruption of the outer parts of the satellite.  Those satellite disc particles that have not yet fallen onto the final remnant are laying on the tidal tails, corresponding to clumps of $\varepsilon$ in the figures.  Material from the satellite bulge shows an inverse trend: initially  in a hot, non-rotating distribution (panel [d]) they end up clustered along the locus of (direct) circular orbits (panels [h] and [l]). We saw in \S\,\ref{Sec:thickening} that bulge particles lie in a flat structure after the merger.  We see now that this structure is a dynamically cold disc.  That satellite bulge particles are deposited into circular orbits in the potential of the remnant indicates that dynamical friction has circularized the merger orbit prior to disruption of the satellite bulge.   Such a disc is too faint for detection if it exists in a real galaxy, except perhaps for inclined views.  However, it may be brighter for other mass ratios, merger configurations, or if gas was present in the satellite that later forms stars \citep[][]{Mihos94}, yielding an inner or nuclear disc not unlike those found in early-type galaxies  \citep[][]{Erwin03,Erwin04}.  Our range of TF exponents yields both inner rings (TF=3: Fig.\,\ref{Fig:energyjz}, panels [g] and [h]) and inner discs (TF=4: Fig.\,\ref{Fig:energyjz}, panels [k] and [l]), as a result of the earlier disruption of lower-density satellites.  

\subsection{Rotation curves of the merger remnants}
\label{Sec:kinematics}

Figure~\ref{Fig:rotcurve1} presents the initial and post-merger rotation curves of the luminous components for all the experiments. These are means of the line-of-sight velocities measured in spatial bins along a virtual slit placed onto the disc major axis. We have used an inclined view with $i=60^\circ$ ($\theta = 60^\circ$, $\phi =90 ^\circ$) instead an edge-on view to enhance the contribution of bulge particles in the center of the rotation curve. The slit length and width are 6.0 and 0.5, respectively. 

The final rotation curves of the total luminous material (solid lines) are quite symmetric, with an overall shape very similar to the rotation curves of the early-type spirals. They all are very similar to the original rotation curve of the primary galaxy; \ie, TF-scaled satellite accretions imprint little changes to the original rotation curve. In the external regions of panels (h) to (k), rotation decreases outwards, as commonly observed in early-types \citep{Casertano91}. In contrast, in the internal regions of the remnants, the primary bulge particles acquire net rotation, making the inner rise steeper than initially. The effect is stronger for denser satellites (compare panels [b] to [d] or [e] to [g]). 

That the inner rotation curve of the primary disc remains largely unmodified during the accretion is an indication that the merger relaxation is gentle in the present models in which satellite densities are set via TF.  The accretion of denser satellites modelled by ABP01 violently steered the inner galaxy.  The inner disc net rotation diminished as random motions increased, while the primary bulge, which eventually suffered a similar-mass merger with the satellite, absorbed much of the remaining orbital angular momentum and ended up rotating even faster than the primary disc (cf.~Fig.\,8 of ABP01).  In contrast, the TF-scaled satellites modelled in the present paper leave the inner disc rotating rapidly, and imprint a weak rotation to the primary bulge.  This rotation is faster for denser satellites of a given mass, which deliver more mass and angular momentum to the remnant center. 

The details of the inner rotation curves are affected by the lack of rotation in the bulges of the initial primary models: the final inner rotation curves would be steeper if the initial bulges had net rotation.  Similarly, the mild counterrotation of the bulges seen in some of the retrograde mergers (e.g., M2R, M3R, see Fig.\,\ref{Fig:rotcurve1}), would disappear if the initial bulges had net rotation.  

Figure~\ref{Fig:rotcurve2} presents the contribution to the final rotation curves of the two luminous components of the satellite separately. The satellite material does not contribute considerably to the final rotation curve in any experiment, due to its low surface brightness. Therefore, although the satellite bulge and disc material counterrotate in all of the retrograde experiments, the central counterrotation can not be detected in the total rotation curves of our retrograde models. This alleviates the problem of  ABP01 high-density models, which showed counterrotation in excess with respect to what it is actually observed \citep[][]{Silchenko97,Pizzella04}.

Dark matter kinematics are also shown in Figs.\,\ref{Fig:rotcurve1} and \ref{Fig:rotcurve2}. Figure \ref{Fig:rotcurve1}  presents the rotation curves for the final dark matter halo, while Fig.\,\ref{Fig:rotcurve2} shows the rotation for the particles belonging to the primary halo and to the satellite halo, separately.  Analyzing dark matter kinematics is beyond the scope of this paper.  We only note that the global dark matter rotation curve exhibits rotation (Fig.\,\ref{Fig:rotcurve1}) only in the inner parts of the models involving more massive satellites.  Initial halos were non-rotating.   Primary dark matter acquires no rotation (Fig.\,\ref{Fig:rotcurve2}), hence the rotation feature seen in the global dark matter rotation is entirely due to (satellite) mass deposition.   

We have tested if our remnants follow the TF relation of the initial satellites and primary galaxy. Asymmetric drift corrections have been applied to the maximum rotational velocities of each remnant viewed edge-on, according to  \citet{Binney87}. They are all offset to lower luminosities than those which should be expected for their $\alpha_\mathrm{TF}$ and $v_\mathrm{circ}$. Direct models exhibit differences of $\sim 0.5$ mag respect to the TF expected value, while retrograde cases present the highest dispersions ($\sim 1.0$ mag). Nevertheless, these values are similar to the observational dispersions of the TF relation in field spirals \citep{Sakai00} and late-type field spirals \citep{Matthews98}, and field and clustered S0's \citep{Neistein99,Hinz01}. Thus, low-satellite accretions could be one of the sources for the dispersion observed in the TF relation.

\subsection{Scaling relations of discs and bulges}
\label{Sec:scaling} 

We now analyze the changes in bulge and disc global parameter relations in our models.  Any dynamical and structural transformation of galaxies must be compatible with the existence of observational scaling relations.  Hence, we enquire, first, whether accretion-driven galaxy transformations show any systematics,  and, second, whether accretion events tend to broaden any pre-existing scaling relations.   We will show here that many, though not all, of the observed scaling relations are compatible with the trends seen in our models.

\begin{figure*}[!]
\centering
\includegraphics[width=12cm]{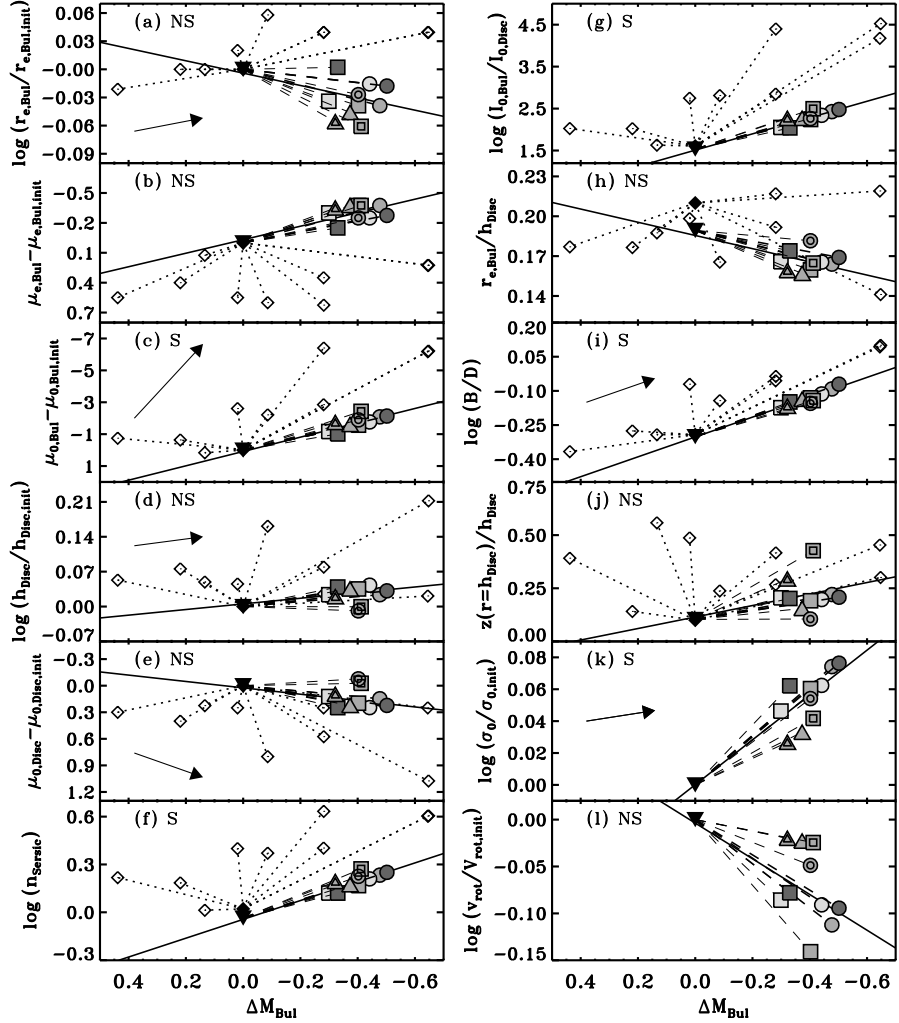}
\caption{The dependence of bulge and disc parameters, and $B/D$ ratio, on the increment of the bulge magnitude (see definition in the text). 
All parameters are derived from Table~\ref{Tab:fits}. Legend is the same as in Fig.\,\ref{Fig:growth}. On the left column, we have plotted photometric parameters obtained directly from the bulge-disc decompositions (scale lengths and central surface densities of bulge and disc, and the S\'ersic index $n$). On the right column, ratios between two of them, disc scale heights and 
typical velocities are shown. \emph{Solid lines}: Orthogonal regressions to the model points. At the right of the identification letter of each panel, "S" indicates that the correlation is significant at a 95\% confidence level, whereas "NS" indicates it is not. \emph{Dashed lines}: Growth vectors of our TF-scaled models. \emph{Dotted lines}: Growth vectors of the high-density models from ABP01. \emph{Arrows}: Slopes of the corresponding observational correlations found by BGP04b. \emph{Panels}: (\emph{a}) Effective radius of the (S\'ersic) bulge component. 
(\emph{b}) Effective surface density of the S\'ersic component. 
(\emph{c}) Central surface density of the S\'ersic 
component, extrapolated from the fit. 
(\emph{d}) Disc scale length. 
(\emph{e}) Extrapolated disc central surface density.
(\emph{f}) S\'ersic index $n$. 
(\emph{g}) Logarithm of the bulge-to-disc ratio of the central intensities. 
(\emph{h}) Ratio between the bulge effective radius and the disc scale length. 
(\emph{i}) Bulge-to-disc luminosity ratio. 
(\emph{j}) Ratio between the disc scale 
height at $r=h_{\mathrm{D}}$, obtained through an exponential fit, and the disc scale length. 
(\emph{k}) Central velocity dispersion. (\emph{l}) Maximum rotational 
velocity of the disc.\label{Fig:parameters}} 
\end{figure*}

We use the photometric parameters in Table~\ref{Tab:fits}. In order to compare with the observed trends, the increments in the photometric parameters of the remnants are plotted vs.\, the increment in the bulge magnitude  in Fig.\,\ref{Fig:parameters}.  Assuming the same $M/L$ for all the bulges \citep{Portinari04}, we have defined the increment in bulge magnitude as
\begin{equation}\label{eq:magnitude}
\Delta M_{\mathrm{Bul}}\equiv M_{\mathrm{Bul}} - M_{\mathrm{Bul,0}}= -2.5\cdot\log\,(\mathcal{M}_{\mathrm{Bul}}/\mathcal{M}_{\mathrm{Bul,0}})
\end{equation}
\noindent where $\mathcal{M}_{\mathrm{Bul}}$ and $\mathcal{M}_{\mathrm{Bul,0}}$ are the masses of the final and initial bulge, respectively. 

Most of the photometric parameters in our models show strong trends with the bulge magnitude increment (Fig.\,\ref{Fig:parameters}). The majority increase towards higher $\mathcal{M}_{\mathrm{Bul}}$, e.g., the S\'ersic index ($n$), the bulge-to-disc ratio ($\log\,(B/D)$), the central and effective bulge surface densities ($\mu _\mathrm{0,Bul}$ and $\mu _\mathrm{e,Bul}$), the disc scale lenght ($h _\mathrm{Disc}$), the bulge-to-disc ratio of the central intensities ($\log\,(I_\mathrm{0,Bul}/I_\mathrm{0,Disc})$), and the central velocity dispersion ($\sigma _\mathrm{0}$); whereas the bulge effective radius ($\log\,(r_\mathrm{e})$), the disc central surface density ($\mu _\mathrm{0,Disc}$), the maximum rotational velocity of the disc ($\log\,(V_\mathrm{rot})$), and the ratio between the bulge effective radius and the disc scale length ($r_\mathrm{e}/h_\mathrm{Disc}$) decrease towards higher $\mathcal{M}_{\mathrm{Bul}}$. Final discs have higher scale lengths and scale heights than the original primary disc, except in two of the retrograde cases (Fig.\,\ref{Fig:parameters},  panels [d] and [j]). Notice that the corresponding relations for the ABP01 dense models (diamonds in Fig.\,\ref{Fig:parameters}) are weaker in all the cases. 

Orthogonal regressions to the distributions are shown in each panel of Fig.\,\ref{Fig:parameters}. The trends exhibited by $\mu _{\mathrm{0,Bul}}$, $n$, $\log\,(I_\mathrm{0,Bul}/I_\mathrm{0,Disc})$, $\log\,(B/D)$ and $\log\,(\sigma _\mathrm{0})$ (panels [(c], [f], [g], [i] and [k] in Fig.\,\ref{Fig:parameters})  have absolute Pearson coefficients above 0.88.  The null correlation hypothesis is rejected for these fits, with a significance of 95\% and for a two-tailed distribution; whereas the rest of correlations are not statistically significant. Nevertheless, the fits involving $\mu _\mathrm{e}$ and the ratio $r_\mathrm{e}/h_\mathrm{Disc}$ exhibit Pearson coefficients above 0.78. The correlations of Fig.\,\ref{Fig:parameters}, together with the correlation found between $\log\,(n)$ and $\log\,(B/D)$ shown in eq.\,\ref{eqn:NvsBD}, indicate that TF-scaled satellite accretion does not alter galaxies randomly: changes in the bulge and disc scaling relations driven by accretion scale systematically with the change in bulge mass.

Many of the changes of photometric parameters from our models follow trends with the bulge magnitude increment similar to the correlations with the bulge magnitude observed by BGP04b. In particular, the disc parameters ($\mu _\mathrm{0,Disc}$ and $h _\mathrm{Disc}$), central velocity dispersions, bulge central surface brightnesses, S\'ersic $n$ and $B/D$ ratios (compare in Fig.\,\ref{Fig:parameters} our correlations -solid lines- with the slopes of the significant observational correlations from BGP04b –-arrows-).  

The trends show some discrepancies with the observations  as well. The bulge effective surface density (Fig.\,\ref{Fig:parameters}b) and the ratio of the central intensities (Fig.\,\ref{Fig:parameters}g) correlate weakly with the bulge magnitude in observations (BGP04b), whereas they do strongly in our models (they exhibit Pearson coefficients $\gtrsim 0.8$). Most relevantly, we fail to reproduce the trends involving the bulge effective radius $r_{\mathrm{e,Bul}}$ (Fig.\,\ref{Fig:parameters}a,h).  $r_{\mathrm{e,Bul}}$ decreases with the accretion events, which is at odds with the fact that the observed more luminous bulges have larger effective radii \citep[][ BGP04b]{Hubble26,Binggeli84,Mollenhoff01}. Such discrepancy diminishes if a higher mass-to-light ratio is assumed for the particles initially belonging to the bulge, while the other correlations do not change significantly. 

\section{Limitations of the present models}
\label{Sec:limitations}

An inherent ingredient of computational work is the strong dependence of the results on the initial conditions.  Our models are no exception, especially given the small coverage of parameter space.  Here we discuss how the specific choices made for our models may have affected the results.  

\begin{enumerate}
\item Using the same initial primary $B/D$ in all the experiments.  

Different initial $B/D$ ratios would likely affect the merger transformation.  
Larger initial $B/D$ would lead to more effective satellite disruption and lower changes in $B/D$.  Conversely, a smaller initial $B/D$ would lead to less efficient disruption of satellites, to stronger non-axisymmetric distortions and more effective inner mixing, which, together, would have led to a faster fractional bulge growth.  This implies that the slope determined for the evolution in the $n$ vs. $B/D$ plane (eqn.\,\ref{eqn:NvsBD}) is valid only for the primaries with $B/D$ similar to those of our models.  Probably, the slope would be higher/lower for smaller/bigger initial $B/D$.  But it is also probable that the slope of the growth in the $n$-$B/D$ plane decreases as the $n$ of the initial primary bulge is higher. This would make the growth mechanism to saturate at $n\sim 3-4$, preventing the departure of the bulges from the observational trend after several consecutive minor mergers. Understanding the behaviour for low-$B/D$ is important as most early- to intermediate-type galaxies have lower $B/D$ than our models (see Fig.\,\ref{Fig:growth}).  
One particularly interesting case is that of bulge-less primary galaxies. Our models suggest that a bulge would easily develop in bulgeless galaxies due to the accretion of even tiny satellites, a manifestation of the well-known problem that bulgeless galaxies pose to hierarchical galaxy formation models \citep{dOnghia04}. The problems for conducting experiments with lower $B/D$ are essentially numerical. Accretion experiments onto smaller bulges require smaller satellites, which imply longer orbital integrations, due to weaker dynamical friction, and smaller time steps, due to shorter dynamical times for bulge and satellite.   

\item The halo profile.

Our haloes have been built with an isothermal profile.  Using different initial halo profiles, e.g. the cosmologically-motivated NFW profile \citep{Navarro95,Navarro97}, might modify the mass deposition, the stripping of the satellite, and the circularization of the orbits. The way in which different halo profiles influence the mass deposition in the remnant center could provide some interesting clues about the central structure of dark haloes.

\item The bulge and halo rotation.

We used non-rotating bulges and halos in the initial models, in order to better constrain the onset of bulge rotation due to the merger.  However, real bulges do rotate \citep[e.g.][]{KormendyIllingworth82}.  Halos probably rotate too --although little information exists on their rotation amplitude or orientation.  Initial models harboring rotating bulges would not only modify the inner parts of the rotation curves, but also the final mass distribution, due to a different spin-orbital coupling.  New models would be required to quantify the strengths of these effects.  

\item The orbit.

We use a single energy and inclination for the merger orbits.  
While most aspects of the merger transformations should remain unchanged under variations in the orbital configurations, differences in the details are to be expected.  Sampling a complete set of preferential orbits according to the distributions extracted from cosmological simulations would allow us to refine our quantitative predictions \citep{ZaritskyRix97,Zaritsky97a,Zaritsky97b,Khochfar06}. 


\item Gas and star formation.

Our models do not include gas hydrodynamics and star formation.  One may argue that the gas must not affect the stellar dynamics, given that gas accounts for less than 10\% of the baryonic mass of a typical spiral galaxy.  Hence, the gas "is there for the ride".  However, changes in the dynamics could occur as gas concentrating in the galaxy nucleus renders the inner potential more centrally-symmetric, and modifies the allowed orbits for stars and gas.  
With or without such effects, the gas would be highly responsive to gravitational perturbations, and any ensuing star formation would alter the stellar make-up of the final bulges and discs.  Changes of $M/L$, perhaps by as much as a factor of $\sim$2  \citep{Portinari04}, would modify our predictions for global scaling relations.  Tests performed by modifying the assumed $M/L$ of bulges and discs, find that the decreasing trend of $\log(r_{\mathrm{e,Bul}})$-$\mathcal{M}_{\mathrm{Bul}}$ (Fig.\,\ref{Fig:parameters}a) reverses sign when increasing the bulge's $M/L$ by a factor of 2.

\item The presence of bars and other galactic components.

Our primary galaxy differs from the vast majority of disc galaxies by not having a central bar \citep[]{Eskridge00,Knapen00}. Initial central components such as bars or lenses, would alter the galactic potencial in the center, and again, the mass distribution might be altered. In particular, a central bar might increase the mixing of bulge and disc material above that given by the merger.  Simulations with bars in the initial models would be required to quantify the effect.  

\item The reference epoch for our primary galaxy model

Because we model our primary galaxy after the MW,  
one may argue that our experiments only test recent bulge growth; and, similarly, that the importance of satellite accretion for the growth of present-day bulges would better be studied using galaxies modelled after those that existed when the minor merger rate was relevant.  
A specific question is whether enough galaxies had well-developed discs for the
processes modeled in this paper to be astrophysically relevant bulge-formation
mechanisms.   One  difficulty is to determine what this reference epoch
might be.  According to CDM cosmological simulations
\citep{Bertschik04a,Bertschik04b} as well as to observations \citep{LeFevre00},
the minor- and major-merger probabilities both peak around  $z\gtrsim 1.0$; although the position of the peak is very dependent on the environment and the mass of object one is looking at \citep{Gottlober01}. At this epoch, the Hubble sequence was in place, galaxy sizes were broadly similar to today's \citep{Simard99,TrujilloAguerri04,Ravindranath04}, and, therefore, our modeling would be applicable.  On the other hand, several lines of evidence point to $1\leq z \leq 2$  as a key epoch for the appearance of present-day early-type galaxian components, i.e., ellipticals and bulges \citep{Lilly99,Eliche06}.  
As we move to higher redshifts, dominant morphologies depart more and more from mature Hubble types.  Galaxies were smaller, gas-rich, less geometrically regular, and were forming stars at higher rates than local galaxies  
\citep{Steidel96,Lowenthal97,Driver98,Bunker00,Trujillo04}.  
Few of the bulges formed at these early cosmic epochs \citep[e.g.][]{Koo05} may have been put in place through the processes modeled here.

\end{enumerate}

\section{Discussion}
\label{Sec:discussion}

We have shown that the accretion of a satellite simultaneously rises $B/D$ and the bulge $n$ index.  More generally, accretion causes a systematic evolution that makes the final structural parameters of the galaxy to be strongly correlated with the bulge luminosity.  Given that the infall of small, collapsed baryonic systems is an inherent ingredient of galaxy formation models based on CDM \citep{KauffmannWhite93,Navarro00}, it is natural to ask what the effect of many such accretion events might be.  Considering that the effects of merging are probably cumulative (ABP01), the lessons derived from the models presented here suggest that bulge growth and $n$ increase would be progressive, leading to a continuous evolution of disc galaxies toward earlier Hubble types.

KK04 list satellite accretion among the "environmental, secular evolution" processes that drive galaxy evolution.  Indeed, as long as satellites galaxies are available, satellite accretion should be recurrent over long timescales.  Even a single accretion event may be termed a secular process given its duration of well over the galaxy's orbital time (our merger models take $\sim$2~Gyr to complete), and the low amplitude of the gravitational perturbation by the satellite.  But KK04 address the role of accretion in (pseudo)bulge growth by focussing on the ability of accretion to furbish gas for star formation.  Our work suggests that accretion events may also contribute to bulge growth through an altogether different process, namely the inward transport of disc stars to the region of the bulge.  

The products of satellite accretion may even share some of the properties that are commonly used to define pseudobulges \citep[as listed by, e.g., ][]{Kormendy05}, and which are usually blamed on disc instabilities. Rapidly-rotating bulges are expected from prograde accretions, which deposit angular momentum in the bulge region and increase any pre-existing rotation (compare before- and after- primary bulge rotation curves in Figure\,\ref{Fig:rotcurve1}).  Furthermore, inward piling-up of disc material in the region of the bulge should add rapidly-rotating material to the region of the photometric bulge.  Inner spiral patterns are also expected from accretions, especially so in low-$B/D$ systems in which gas from the primary or from the satellite settle into an inner disc. 
The limited resolution of our models prevents us from making quantitative statements.  Measurement of a bulge rotation velocity not affected by the inner disc is hard given the thickness of the disc.  Measuring an ellipticity for the bulge is hard for the same reason.  

X-/peanut-shaped structures, another definitory feature for pseudo-bulges, are not seen in any of our models.  However, \citet{Mihos95} show an accretion model which results in a strong X-shaped structure.  Their primary galaxy, which lacks a bulge component, develops a strong bar which buckles into a peanut-shape.   In our models, the presence of a massive bulge in the primary galaxy stabilizes the disc against a strong bar response \citep[][GGB05]{Sellwood99,Debattista05}.  Peanut-shaped bulges may thus appear after the accretion into low-$B/D$ galaxies.  

The above arguments suggest that satellite accretion may play a role in setting those signatures commonly used to define pseudobulges.  The role of the accretion is that of a trigger of the disc perturbation which leads to the infall of disc material to feed the bulge.

In fact, satellite accretion events can account for several observational facts that disc internal instabilities can not. Firstly, they can connect external and internal processes, as several works have addressed that bars are easily excited by tidal interactions \citep[][]{Noguchi87,Gerin90,Miwa98}. And secondly, minor mergers are in agreement with the statistical association of young populations with morphological disturbances reported by \citet{Kannappan04}; while being, at the same time, also compatible with the low fraction of very peculiar/distorted shapes found at $z\lesssim$0.4 \citep[$\lesssim$10\%, see][]{vandenBergh00}. Satellite accretion events can be the key processes to reconcile the high merging rates of hierarchical formation scenarios and the fact that many disc galaxies do not exhibit signs of major merger violence.

\section{Conclusions}
\label{Sec:conclusions}

Our simulations of TF-scaled satellite accretion experiments show that:

\begin{enumerate} 

\item The TF-scaled satellite accretion drives pre-existing discs to a secular evolution towards earlier types (higher $B/D$ and $n$). More massive satellites produce higher changes in $n$, $B/D$ ratio, and disc thickening. 

\item  For the $B/D$=0.5 primaries modeled here, $B/D$ and $n$ grow such that,  roughly, $n$ increases by a factor of $\sim 2.3$ when $B/D$ doubles.  Steeper growth of $n$ is expected in primary galaxies with smaller initial bulges.  

\item The evolution of the bulge component is driven foremost by the injection of disc material to the center through transitory non-axisymmetric distortions, and to a lesser extent by expansion of the pre-existing bulge and by deposition of satellite material.  Vertical heating thickens the disc.  

\item The tidal radius of the satellite depends critically on the choice of TF exponent assumed for the scaling of galaxy densities.  For the $B/D$=0.5 primaries modeled here, $\alphaTF = 3$ leads to complete disruption and to a torus of accreted material, while the satellites reach the center for $\alphaTF= 4$.  

\item  The satellite disc deposits its mass onto the remaining primary disc. Particles from the satellite bulge settle into a rather cold inner disc or ring, due to orbit circularization.  

\item The radial and vertical redistribution of material by heating produces a population mixture.  Part of the material initially belonging to the primary disc is interpreted as bulge in the decompositions. The photometrically identified final bulge is partially supported by rotation, as it is observed in several galaxies.

\item Final rotation curves exhibit symmetric shapes, and a gentle outward decline typical of early-type disc galaxies.  
The satellite material does not dominate the central parts of the final remnants in any case, and hence, central counterrotation is not visible in the total rotation curves of retrograde mergers. 

\item Accreted TF-scaled satellites cause systematic structural transformations in the primary galaxy, which evolves towards higher $B/D$, higher $n$, higher $\sigma _{\mathrm{0}}$, higher $h_{\mathrm{Disc}}$ and lower $\mu _{\mathrm{0,Disc}}$, all following trends with the bulge magnitude similar to the  correlations observed by BGP04b. On the other hand, some observational trends that are observed to not correlate with the bulge magnitude do it strongly in our remnants, as the ratio between the central intensities of the bulge and the disc, and the bulge effective surface density. We fail to reproduce the trends involving the bulge effective radius $r_{\mathrm{e,Bul}}$. 

\item Secular evolution of galaxy bulges, currently being discussed as the outcome of bar instabilities in the disc due to gas cooling, can be due to satellite accretion as well, through the inward transport of disc material to the region of the bulge. 

\end{enumerate}

\begin{acknowledgements}
We wish to acknowledge K.~ Kuijken and J. Dubinski, and L. Hernquist for making their galaxy generating code and his version of the {\small TREECODE\/} available.  We thank Mercedes Prieto and Reynier Peletier  for interesting discussions and comments on the paper, and the referee  for suggestions and comments that helped to improve the manuscript.
\end{acknowledgements}

\bibliographystyle{aa}

\end{document}